\documentclass[showpacs,nofootinbib,preprintnumbers,prd,superscriptaddress,onecolumn]{revtex4}
\usepackage{adjustbox}
\usepackage{amsmath}
\usepackage{amsfonts}
\usepackage{amssymb}
\usepackage{graphicx}
\usepackage[titletoc]{appendix}
\usepackage{color}
\usepackage{hyperref}
\usepackage{cleveref}
\usepackage[rightcaption]{sidecap}
\usepackage{subfigure}
\usepackage{comment}
\usepackage{tensor}
\usepackage{longtable}
\usepackage{graphicx}

\usepackage{mathtools}

\usepackage{dcolumn}

\usepackage{array}
\usepackage{ctable}
\usepackage{multirow}
\usepackage{siunitx}
\usepackage{longtable}
\usepackage{tabularx}
\usepackage{booktabs}

\usepackage{longtable}
\graphicspath{{Graphics/}}

\def\be{\begin{equation}}
\def\ee{\end{equation}}
\def\bea{\begin{eqnarray}}
\def\eea{\end{eqnarray}}

\definecolor{vividviolet}{rgb}{0.62, 0.0, 1.0}
\definecolor{amaranth}{rgb}{0.9, 0.17, 0.31}
\definecolor{palatinateblue}{rgb}{0.15, 0.23, 0.89}
\definecolor{brightpink}{rgb}{1.0, 0.0, 0.5}
\definecolor{cornflowerblue}{rgb}{0.39, 0.58, 0.93}
\definecolor{deepcarminepink}{rgb}{0.94, 0.19, 0.22}
\definecolor{radicalred}{rgb}{1.0, 0.21, 0.37}

\hypersetup{ linktoc=all,
    colorlinks, linkcolor={palatinateblue},
    citecolor={brightpink}, urlcolor={amaranth}
}

\begin{document}

\title{Stability analysis of dilaton-inspired scalar field within the geometrical trinity of gravity}

\author{Youri Carloni}
\email{youri.carloni@unicam.it}
\affiliation{Universit\`a di Camerino, Via Madonna delle Carceri, Camerino, 62032, Italy.}
\affiliation{INAF - Osservatorio Astronomico di Brera, Milano, Italy.}
\affiliation{Istituto Nazionale di Fisica Nucleare (INFN), Sezione di Perugia, Perugia, 06123, Italy.}

\author{Orlando Luongo}
\email{orlando.luongo@unicam.it}
\affiliation{Universit\`a di Camerino, Via Madonna delle Carceri, Camerino, 62032, Italy.}
\affiliation{INAF - Osservatorio Astronomico di Brera, Milano, Italy.}
\affiliation{SUNY Polytechnic Institute, 13502 Utica, New York, USA.}
\affiliation{Istituto Nazionale di Fisica Nucleare (INFN), Sezione di Perugia, Perugia, 06123, Italy.}
\affiliation{Al-Farabi kazakh National University, Al-Farabi av. 71, 050040 Almaty, kazakhstan.}

\author{Andronikos Paliathanasis}
\email{anpaliat@phys.uoa.gr}
\affiliation{Department of Mathematics, Faculty of Applied Sciences, Durban University of Technology, Durban 4000, South Africa}
\affiliation{School for Data Science and Computational Thinking and Department of
Mathematical Sciences, Stellenbosch University, Stellenbosch, 7602, South
Africa}
\affiliation{Departamento de Matemáticas, Universidad Católica del Norte,
Avda. Angamos 0610, Casilla 1280 Antofagasta, Chile}

\begin{abstract}
We investigate the dynamics of the dilaton-inspired scalar field, formally rewritten by means of a Brans-Dicke Lagrangian, within the framework of \emph{geometrical trinity of gravity}. In this respect, we perform a stability analysis by adopting a non-flat Friedmann-Robertson-Walker (FRW) metric and considering the well-established exponential potential in three distinct gravitational frameworks: general relativity, teleparallel gravity, and symmetric-teleparallel gravity. By comparing the scalar field behaviors across these theories, we highlight the role of curvature, torsion, and non-metricity in shaping cosmic evolution. Our analysis reveals that, both in general relativity and teleparallel gravity, the dilaton-inspired field can drive the accelerated expansion of the universe, effectively behaving as cosmological constant at late times. In contrast, within the symmetric teleparallel gravity scenario, performing a complete linear stability analysis is prevented by the use of the non-coincident gauge. Nevertheless, the latter paradigm introduces complexity into the autonomous system, resulting in a structurally different analysis. For general relativity and teleparallel scenarios, we remark the regions of attractor solutions and unphysical domains in which we do not expect the viability of our dilaton-inspired Lagrangian. However, within the framework of symmetric-teleparallel gravity, the stability analysis reveals no attractor points for the chosen set of free parameters. In support of these findings, physical conclusions, kinematical studies, and consequences on Friedmann dynamics are thus explored.
\end{abstract}


\maketitle
\tableofcontents

\section{Introduction}

The current accelerated expansion of the universe is well explained by the $\Lambda$CDM model, where $\Lambda$ indicates the cosmological constant and CDM denotes the cold dark matter \cite{SupernovaCosmologyProject:1997zqe,SupernovaSearchTeam:1998fmf,SupernovaCosmologyProject:1998vns,WMAP:2003elm,SupernovaSearchTeam:2003cyd}. Although it successfully explains most of the observational data, the cosmological constant problem \cite{Weinberg:2000yb,Carroll:2000fy,Dolgov:1997za,Padmanabhan:2002ji,Martin:2012bt} and persistent tensions \cite{DiValentino:2021izs,Giare:2024akf,Poulin:2022sgp,Joseph:2022jsf} motivate the exploration of dynamical dark energy or extended scenarios \cite{Aviles:2011sfa} as alternative.

Recent analyses from DESI collaboration  \cite{DESI:2024mwx,DESI:2024aqx,Colgain:2024xqj,Cortes:2024lgw,Shlivko:2024llw,Giare:2024gpk,Carloni:2024zpl,Luongo:2024fww,DESI:2025kuo,DESI:2025wyn,Cortes:2025joz,Dinda:2025svh,Colgain:2025nzf,Paliathanasis:2025hjw} support this viewpoint, identifying the $w_0w_a$ parametrization as a more accurate representation of late time cosmic acceleration. In this context, scalar fields emerge as fundamental components in modeling dynamical dark energy \cite{Steinhardt:1999nw,Chiba:2002mw,Chiba:2005tj,Scherrer:2007pu,Scherrer:2008be}.

In other words, the role played by scalar fields revived in the context of cosmology, even at late times, while being contextual at early stages, where inflation is thought to be driven by virtue of the inflaton.

A non-exhaustive list of possibilities is offered by

\begin{itemize}
    \item[-] scalars in analogy to quintessence fields \cite{Kamenshchik:2001cp,Copeland:2006wr,Amendola:1999er,Chiba:1999ka},
    \item[-] K-essence models and/or paradigms with negative kinetic terms \cite{Copeland:2006wr,Scherrer:2004au,Scherrer:2008be,Luongo:2025iqq},
    \item[-] generalized K-essence models, resembling matter, under the form of quasi-quintessence \cite{Luongo:2024opv,Carloni:2025cjj,Luongo:2018lgy,DAgostino:2022fcx},
    \item[-] Higgs fields applied to cosmological contexts, such as inflation \cite{Bezrukov:2008dt,Bezrukov:2010jz,Rubio:2018ogq},
    \item[-] boson condensate and/or mixing of scalar fields, under the form of particle-like configurations \cite{Goodman:2000tg,Das:2014agf},
    \item[-] dilaton, inspired by string theory or low-energetic domains of superstrings \cite{Tseytlin:1991xk,Tseytlin:1991ss},
\end{itemize}
and so on, see e.g. \cite{Belfiglio:2023rxb,Ferreira:1997hj,Poulin:2018cxd,Nojiri:2005pu,Lim:2010yk,Padmanabhan:2002cp,Dunsby:2016lkw,Capozziello:2018hly}.

In this work, we investigate a dilaton-inspired scalar field as a candidate for dynamical dark energy \cite{Bean:2000zm}. This scenario is derived by considering the Brans-Dicke theory in the Jordan frame, which mimics a low-energy effective superstring action for the graviton and dilaton scalar field \cite{Brustein:1994kw,Cartier:1999vk,Dabrowski:2005yn}.

A distinctive feature of the dilaton action is its invariance under a duality transformation \cite{Buscher:1987sk,Alvarez:1994wj}.
In the absence of spatial curvature, the corresponding cosmological field equations exhibit scale factor duality symmetry, as formulated by the Gasperini–Veneziano transformation \cite{Gasperini:1992em}. This duality, founded upon the $O(d,d)$ symmetry of the gravitational action \cite{Buscher:1987sk,Alvarez:1994wj,kehagias:1996st}, sheds light on the pre-big bang phase and enables the construction of nonsingular, nonperturbative cosmological models \cite{Gasperini:1992em,Bossard:2002ta,Brustein:1998kq}. Although this symmetry is preserved under conformal transformations, the physical interpretation of the duality is generally lost in the transformed frame \cite{Gionti:2017ffe}. Generalizations of this property have been explored in the context of the teleparallel dilaton action in Refs. \cite{Paliathanasis:2021gfq,Paliathanasis:2024tit}. Furthermore, the $O(d,d)$ symmetry plays a significant role in the Wheeler–DeWitt formulation of quantum cosmology \cite{kehagias:1996st}.

In view of these results, we extend the Brans–Dicke theory beyond curvature-based gravity to incorporate torsion and non-metricity scalars. Previous studies have examined the dilaton action in a spatially flat FRW geometry, including curvature-based scenarios \cite{Hrycyna:2013yia}, torsion-based formulations \cite{Paliathanasis:2021nqa}, and non-metricity frameworks \cite{Paliathanasis:2024tit}. In this paper, we focus on cosmologies with non-zero spatial curvature in the FRW metric, consistent with Planck observational data \cite{Planck:2018vyg}. The impact of spatial curvature, in conjunction with non-metricity, is explored in the coincident gauge in Ref. \cite{Jensko:2024bee}, establishing a link between $f(Q)$ \cite{BeltranJimenez:2019tme,Heisenberg:2023lru} and $f(T)$ \cite{Ferraro:2006jd} gravity theories in a non-flat FRW scenario.

In addition, curved inflationary models have been investigated recently, and it has been shown that inflation is unaffected by negative curvature, while the energy density associated with the curvature can be non-zero during the pre-inflationary era. Then, by applying the cosmological principle, it is possible to realize a homogeneous and isotropic open or closed inflationary scenario that evolves toward a spatially flat universe at late times \cite{Steigman:1983hr,Mathews:2014fwa,Aslanyan:2015pma,Paliathanasis:2016dhu}.

Motivated by these considerations, we examine the behavior of our scalar field within non-metricity gravity, adopting the non-coincident gauge to enable a comprehensive comparison of the differences between non-metricity and torsion backgrounds. In particular, we use the same dimensionless variables to construct autonomous systems for both curvature and torsion backgrounds, assuming an open or closed universe \emph{a priori}. By contrast, a distinct approach is employed in the non-metricity scenario: the non-coincident gauge introduces an additional scalar field, which requires spatial curvature to be treated as a dynamical variable. Consequently, this demands a different set of dimensionless variables, modifying the stability analysis. We then perform a detailed phase-space analysis based on an exponential potential, allowing us to reconstruct the cosmological evolution and determine the viability of the model \cite{Copeland:2009be,Tot:2022dpr,Leon:2012vt,Lazkoz:2007mx,Paliathanasis:2020sfe,Carloni:2024ybx,Bajardi:2022tzn,Carloni:2023egi}. This approach also allows us to constrain the free parameters to ensure the cosmological viability of the theory \cite{Amendola:2006we, Setare:2012ry}. In this way, we effectively analyze the effects of the dilaton-inspired scalar field within the framework of the \emph{geometrical trinity of gravity} \cite{BeltranJimenez:2019esp}, emphasizing the distinctions among the Ricci scalar $R$, the torsion scalar $T$, and the non-metricity scalar $Q$.

The structure of the paper is as follows: Sect. \ref{sec2} outlines the theoretical framework for the dilaton cosmology. In Sect. \ref{sec3}, we introduce the autonomous systems in a non-flat FRW metric and derive the critical points for the $R$, $T$, and $Q$ theories of gravity. Then, in Sect. \ref{sec4}, we perform a stability analysis and present the phase-space graphs in \emph{geometrical trinity of gravity}. Finally, in Sect. \ref{sec5}, we provide our conclusions and outlook.


\section{Theoretical setup}\label{sec2}

Dilatons are string theory-inspired fields, whose use is particularly important in several contexts \cite{Wetterich:1987fm, Tseytlin:1992ye,Kostouki:2009sk,Garcia-Bellido:2011kqb,Campbell:2011iw}. For example, from dilaton couplings is possible to formulate a particle nature of black holes \cite{Horne:1992zy,Horne:1992bi,Alves:2000pe,Mignemi:1992pm} or to define spacetime regular solutions, analogously to non-linear electrodynamics \cite{Gibbons:1995ap,Berman:1997iz,Gibbons:2000mx,Unver:2010uw,Chan:1995fc} or, in principle, to ensure a particle-like behavior to spacetimes \cite{Duff:1993ye,Porfyriadis:2023qqo,Burgess:1994kq} and so on, see e.g. \cite{Paliathanasis:2021gfq,Leon:2018skk,Chu:2015ila,Ellis:2014cma,Bellazzini:2013fga,Bezrukov:2012hx}.

Non-minimally coupling dilatons to gravity may be also suggested by Mach's principle \cite{Brans:1961sx,Darabi:2008duw,Barbour:2002ad} and turns out to be important in contexts such as dark matter \cite{Dick:1996aw,Cho:1998js,Cho:2007cy,Susperregi:2003gx}, dark energy \cite{Huang:2010xr,Catena:2007jf,Susperregi:2003gx,Piazza:2004df} or, in general, in field theories.

At the same time, one can imagine to \emph{extend} the idea of gravity under the form of pure geometry, including alternatives such as torsion and/or nonmetricity \cite{Hehl:1976kj,Hehl:1994ue}. Alternatively speaking, if Einstein formulated his theory in view of torsion, for instance, it would be possible to reproduce almost same outcomes by adopting a further spin, that however has not been detected yet \cite{Hayashi:1979qx,Gomes:2023hyk}.

The interest in compelling theories of gravity that appear analogous to curvature appears particularly evident in recent year, while suggesting the concept of \emph{trinity of gravity} \cite{BeltranJimenez:2019esp}.

Hence, it appears quite interesting to explore the potential consequences of introducing a dilaton-inspired field in contexts of gravity that may be, or may depart from, curvature.

This is the case in the \emph{geometrical trinity of gravity}, which aims to distinguish the influence of such a field across different gravitational backgrounds.

Thus, it would be intriguing to consider three different versions of the gravitational actions for the scalar-tensor, scalar-torsion, and scalar-nonmetricity theories \cite{Faraoni:2004pi,Hohmann:2018dqh,Gonzalez-Espinoza:2019ajd,Paliathanasis:2023gfq,Jarv:2023sbp} coupled with dilatons, to both identify\footnote{Throughout this work, we adopt the metric signature $(-,+,+,+)$ and set $8\pi G = c = 1$.}:
\begin{itemize}
    \item[-] departures from the geometric interpretation of gravity in presence of dilaton fields,
    \item[-] stability of the underlying models, in view of the Mach's principle recipe applied to Lagrangians, far from the Hilbert-Einstein action.
\end{itemize}

To do so, consider the following set of actions,

\begin{align}
&\mathcal{S}_R=\int d^{4}x\sqrt{-g}\left(\frac{F(\varphi)}{2}R-\frac{\omega(\varphi)}{2}g^{\mu\nu}\varphi_{,\mu}\varphi_{,\nu}-V\left(\varphi\right)\right),\label{eq:SR}\\
&\mathcal{S}_T=\int d^{4}x\sqrt{-g}\left(\frac{F(\varphi)}{2}T-\frac{\omega(\varphi)}{2}g^{\mu\nu}\varphi_{,\mu}\varphi_{,\nu}-V\left(\varphi\right)\right),\label{eq:ST}\\
&\mathcal{S}_Q=\int d^{4}x\sqrt{-g}\left(\frac{F(\varphi)}{2}Q-\frac{\omega(\varphi)}{2}g^{\mu\nu}\varphi_{,\mu}\varphi_{,\nu}-V\left(\varphi\right)\right),\label{eq:SQ}
\end{align}
which can be summarized by considering the set $\Upsilon\equiv \{R;T;Q\}$, selecting the typology of gravity interaction\footnote{The set $\Upsilon$ is neither a vector or a particular choice, but simply a way to compactly describe the three distinct Lagrangian formalisms developed in Eqs.~\eqref{eq:SR},\eqref{eq:ST}, and \eqref{eq:SQ}, i.e., appearing quite easy in its use for our later purposes.}, constructing \emph{de facto} the following general action
\begin{equation}
\mathcal{S}_{\Upsilon}=\int d^{4}x\sqrt{-g}\left(\frac{F(\varphi)}{2}\Upsilon-\frac{\omega(\varphi)}{2}g^{\mu\nu}\varphi_{,\mu}\varphi_{,\nu}-V\left(\varphi\right)\right),\label{eq:genact}
\end{equation}
where $\Upsilon$ represents one of the geometric scalars that constitute the trinity of gravity, i.e., the Ricci scalar $R$ when we consider the Levi-Civita connection, the torsion scalar $T$ in teleparallel gravity, and the non-metricity scalar $Q$ in symmetric-teleparallel gravity.

The functional forms of the above relations, Eqs.~\eqref{eq:SR},\eqref{eq:ST} and \eqref{eq:SQ}, appear quite complicated to explore in terms of stability. Accordingly, we can reformulate the actions by observing that Eq.~\eqref{eq:genact} can be rewritten in the perspective of a generic Brans-Dicke model, through the identifications  $F(\varphi)=\varphi$, $\omega(\varphi)=\frac{\omega_{0}}{\varphi}$, with $\omega_{0}=\text{const.}$, furnishing
\begin{equation}
\mathcal{S}_{\Upsilon}^{BD}=\int d^{4}x\sqrt{-g}\left(\frac{\varphi}{2}\Upsilon-\frac{\omega_0}{2\varphi}g^{\mu\nu}\varphi_{,\mu}\varphi_{,\nu}-V\left(\varphi\right)\right).
\end{equation}
This intermediate step can be rewritten in an artful way, by redefining the scalar field as $\varphi = e^{\phi}$ and working out the exponential potential, $\hat{V}(\phi) = V(\phi)e^{-\phi}$. Then, the above actions turn out to be
\begin{equation}
\mathcal{S}^{D}_{\Upsilon}= \int d^{4}x\sqrt{-g} e^{\phi}\left(\frac{\Upsilon}{2} - \frac{\omega_{0}}{2} g^{\mu\nu} \phi_{,\mu} \phi_{,\nu} - \hat{V}(\phi) \right) ,\label{eq:SDIL}
\end{equation}
that provide a viable dilaton-inspired form as shown in Refs. \cite{Brustein:1994kw,Cartier:1999vk,Dabrowski:2005yn,Paliathanasis:2023gfq}, while being particularly practical and manageable in computing non-dimensional variables through which we can argue the dynamical behaviors of Eqs.~\eqref{eq:genact}.

The main differences with a genuine dilaton fields consist in the absence of vector fields, non-minimally coupled with $\phi$ or $\varphi$, plus the trick to obtain it from a generic, and more complicated, non-minimal Lagrangian, identified by the field, $\varphi$, in Eqs.~\eqref{eq:SR},\eqref{eq:ST} and \eqref{eq:SQ}. Hereafter, the terms dilaton-inspired and dilaton will be used interchangeably and, therefore, we pursue the following conceptual and physical steps.

\begin{itemize}
  \item[-] We examine the behavior of the dilaton field through an effective representation based on the Brans-Dicke scalar-tensor formalism.
  \item[-] We focus, as appropriate in each instance, singling out the Ricci scalar, torsion, and non-metricity frameworks.
  \item[-] The study focuses on stability, applied to the universe, namely invoking the \emph{cosmological principle}, aiming at applications from small to larger scales.
  \item[-] As the dilaton field is specifically considered within the low-energy effective superstring action, we may imagine to makes the model applicable to early-universe scenarios, where spatial curvature cannot be neglected \cite{Alves:2000pe,Paliathanasis:2021gfq,Wetterich:1987fm} and, therefore, we will model the non-dimensional variables through it.
\end{itemize}

Accordingly, we thus investigate the dilaton scalar field within the trinity of gravity using the FRW line element, employing  non-zero spatial curvature
\begin{equation}
ds^{2} = -N^{2}(t) \, dt^{2} + a^{2}(t) \left(\frac{dr^{2}}{1 - k r^{2}} + r^{2} d\Omega^{2}\right),
\end{equation}
where $N(t)$ is the lapse function, $a(t)$ is the scale factor, and $d\Omega^{2} \equiv d\theta^{2} + \sin^{2}\theta \, d\varphi^{2}$ represents the angular component. The parameter $k = \pm 1$ denotes the spatial curvature of the three-dimensional hypersurface, corresponding to a closed or open universe. Here, the lapse function is introduced to analyze the minisuperspace description of the gravitational theories. When a minisuperspace model is considered, the gravitational field equations can be significantly simplified, allowing certain degrees of freedom to be described equivalently by a scalar field \cite{Ryan:1975jw,Paliathanasis:2023pqp}.

To better shed light on the dilaton behavior, we involve an action characterized by the dilaton dominance over the other species and, meanwhile, we assess the viability of the scenario in terms of background stability, within one scalar inside the set $\Upsilon$, defined in Eq.~\eqref{eq:genact}.

To do so, let us now summarize the mathematical structures associated with each underlying theory, namely considering Einstein's gravity, torsion and non-metricity.

\subsection{Scalar field representation in general relativity}

In general relativity, gravity is represented by the Ricci scalar, which, when adopting the FRW metric, is expressed as \cite{Paliathanasis:2025sft}

\begin{equation}
R = 6\left(  \frac{1}{N^{2}}\frac{\ddot{a}}{a} + \left(  \frac{1}{N}\frac{\dot{a}}{a}\right)^{2} - \frac{\dot{N}}{N^{3}}\frac{\dot{a}}{a} + \frac{k}{a^{2}} \right),
\end{equation}
from which the point-like Lagrangian for the dilaton scalar field is derived as
\begin{equation}
\mathcal{L}_{R}\left(  N,a,\dot{a},\phi,\dot{\phi}\right) = e^{\phi}\left(  -\frac{6}{N}a\dot{a}^{2} - \frac{6}{N}a^{2}\dot{a}\dot{\phi} - \frac{\omega_{0}}{2N}a^{3}\dot{\phi}^{2} + Na^{3}\left( V\left(  \phi\right) + 6\frac{k}{a^{2}} \right)  \right). \label{eq:LR}
\end{equation}
At this stage, by varying Eq.~\eqref{eq:LR} with respect to $\{N,a,\phi\}$ and by setting the lapse function $N=1$, we derive the Friedmann's equations and the Klein-Gordon modified equation as follows
\begin{subequations}
\begin{align}
    &6Hk a^{-2}+6 H \dot{\phi}+6 H^2+V(\phi)+\frac{1}{2} \omega_{0} \dot{\phi}^2=0\label{eq:FRWRN},\\
    &12 \dot{H}+6 k a^{-2}+12 H \dot{\phi}+18 H^2+3 V(\phi)+6 \ddot{\phi}-\frac{3}{2} \omega_{0} \dot{\phi}^2+6 \dot{\phi}^2=0\label{eq:FRWRa},\\
     &\omega_{0}\ddot{\phi}+\frac{1}{2} \omega_{0}\dot{\phi}^2+\dot{V}(\phi)+V(\phi)+6 \dot{H}+6 k a^{-2}+3 \omega_{0} H\dot{\phi}+12 H^2=0\label{eq:FRWRphi},
\end{align}
\end{subequations}
where dot denotes the derivatives with respect to the cosmic time $t$ and $H\equiv \frac{\dot{a}(t)}{a(t)}$ is the Hubble function.

\subsection{Scalar field representation in
 teleparallel theory}

In teleparallel theory, metric $g_{\mu\nu}$ in terms of the vierbein field $e^A_\mu$ is provided by $g_{\mu\nu} =  e^A_\mu e^B_\nu\eta_{AB}$,
where $\eta_{AB} = \text{diag} (-, +, +, +)$.

In this scenario, the spin connection $\omega^A_{\ B\mu}$ determines the parallel transportation, then the torsion tensor and the corresponding contorsion tensor are given by
\begin{align}
    &T^A_{\ \mu\nu}(e^A_{\ \mu},\omega^A_{\ B\mu})=
\partial_\mu e^A_{\ \nu} -\partial_\nu e^A_{\ \mu}+\omega^A_{\ B\mu}e^B_{\ \nu}
-\omega^A_{\ B\nu}e^B_{\ \mu},\\
    &k^{\mu\nu}_{\  \ A}=\frac{1}{2}
\left(T^{\ \mu\nu}_{A}
+T^{\nu\mu}_{\ \ A}
-T^{\mu\nu}_{\  \ A}\right).
\end{align}

Accordingly, the torsion scalar,
$T$, becomes
$T= T^A_{\ \mu\nu}S_A^{\ \mu\nu}$, in which
    $S_A^{\ \mu\nu}= e_A^{\ \rho}S_\rho^{\
\mu\nu}=k^{\mu\nu}_{\ \ A}
-e^A_{\ \nu}T^{ \mu}
+e^A_{\ \mu}T^{ \nu}$, where the vierbein field $e^A_\mu = \left(-N(t), \frac{a(t)^2}{1-kr^{2}}, a(t) r^{2}, a(t)^{2}r^{2}\sin^{2}\theta\right)$ is expressed in the Weitzenböck gauge with vanishing spin connection\footnote{This choice is the only one guaranteeing both the vierbein and the teleparallel connection are compatible with the cosmological symmetries of a non-flat FRW metric \cite{Bahamonde:2022ohm,Carloni:2025cjj}.}.

With these considerations, we obtain \cite{Paliathanasis:2021uvd}
\begin{equation}
T = 6\left( \frac{k}{a^{2}} - \frac{1}{N^{2}}\left(  \frac{\dot{a}}{a}\right)^{2} \right),\label{eq:Tscalar}
\end{equation}
which, when substituted into the action in Eq.~\eqref{eq:SDIL}, leads to the point-like Lagrangian
\begin{equation}
\mathcal{L}_{T}\left(  N,a,\dot{a},\phi,\dot{\phi}\right) = e^{\phi}\left(  -\frac{6}{N}a\dot{a}^{2} - \frac{\omega_{0}}{2N}a^{3}\dot{\phi}^{2} + Na^{3}\left(  V\left(  \phi\right) + 6\frac{k}{a^{2}} \right)  \right). \label{eq:LT}
\end{equation}
Here, the cosmological field equations, again obtained by varying Eq.~\eqref{eq:LT} with respect to $\{N, a, \phi\}$ and considering $N=1$, are provided by
\begin{subequations}
\begin{align}
    &6 k a^{-2}+6 H^2+V(\phi)+\frac{1}{2} \omega_0 \dot{\phi}^2=0\label{eq:FRWTN},\\
    &12 \dot{H}+6 k a^{-2}+12 H\dot{\phi}+18 H^2+3 V(\phi)-\frac{3}{2} \omega_0 \dot{\phi}^2=0\label{eq:FRWTa},\\
     &\omega_0 \ddot{\phi}+\frac{1}{2} \omega_0 \dot{\phi}^2+\dot{V}(\phi)+V(\phi)+6 k a^{-2}+3 \omega_0 H \dot{\phi}-6 H^2=0\label{eq:FRWTphi}.
\end{align}
\end{subequations}

\subsection{Scalar field representation in
 symmetric-teleparallel theory}

In the symmetric-teleparallel  both the Ricci and torsion tensors vanish, and gravity is determined only by the non-metricity tensor, i.e.,
\begin{equation}
    Q_{\lambda \mu \nu}=\nabla_{\lambda}g_{\mu\nu}=\frac{\partial g_{\mu\nu}}{\partial x^{\lambda}}-\Gamma^{\sigma}_{\ \lambda\mu}g_{\sigma\nu}-\Gamma^{\sigma}_{\ \lambda\nu} g_{\sigma\mu},
\end{equation}
where $\Gamma^{k}_{\ \mu\nu}$ denotes the general affine connection \cite{Mandal:2020lyq,Carloni:2023egi,koussour:2023hgl}. By using the non-metricity tensor, we obtain the non-metricity scalar
\begin{equation}
    Q=Q_{\lambda\mu\nu}P^{\lambda\mu\nu},
\end{equation}
where $P^{\lambda\mu\nu}$ is the superpotential, and it is explicit written as
\begin{equation}
    P^{\lambda}_{\ \mu\nu}=-\frac{1}{4}Q^{\lambda}_{\ \mu\nu}+\frac{1}{2}Q_{(\mu \ \nu)}^{\ \ \lambda}+\frac{1}{4}\left(Q^{\lambda}-\bar{Q}^{\lambda}\right)g_{\mu\nu}-\frac{1}{4}\delta^{\lambda}_{(\mu}Q_{\nu)}.
\end{equation}

Here, $Q_{\mu} = Q_{\mu\nu}^{\ \ \nu}$ and $\bar{Q} = Q^{\nu}_{\ \ \mu\nu}$ represent traces of the non-metricity tensor, while parentheses around the indices denote symmetrization.

Teleparallel theories feature a special gauge in which the connection trivializes and vanishes. In teleparallel geometry, it corresponds to the Weitzenböck gauge, and in symmetric-teleparallel geometry, this is called the coincident gauge.

However, the coincident gauge is criticized due to the presence of ghosts' degrees of freedom \cite{Gomes:2023tur,Heisenberg:2023wgk}. For this reason, we determine the non-metricity scalar in the FRW metric in the non-coincident gauge as \cite{Paliathanasis:2023pqp}

\begin{equation}
Q = -\frac{6\dot{a}^{2}}{N^{2}a^{2}} + \frac{3\gamma}{a^{2}}\left(  \frac{\dot{a}}{a} + \frac{\dot{N}}{N}\right) + \frac{3\dot{\gamma}}{a^{2}} + k\left[  \frac{6}{a^{2}} + \frac{3}{\gamma N^{2}}\left(  \frac{\dot{N}}{N} + \frac{\dot{\gamma}}{\gamma} - \frac{3\dot{a}}{a}\right)  \right].\label{eq:Qscalar}
\end{equation}

In this case, the component $\gamma$ can be physically interpreted as a new field entering the expression of the symmetric-teleparallel theory. Then, by substituting $\gamma = \frac{1}{\dot{\Psi}}$, the point-like Lagrangian can be written as
\begin{equation}
\mathcal{L}_{Q}\left(  N,a,\dot{a},\phi,\dot{\phi},\Psi,\dot{\Psi}\right) = e^{\phi}\left(  -\frac{6}{N}a\dot{a}^{2} - \frac{\omega_{0}}{2N}a^{3}\dot{\phi}^{2} - 6Na\frac{\dot{\phi}}{\dot{\Psi}} + \frac{6}{N}ka^{3}\dot{\phi}\dot{\Psi} + Na^{3}\left(  V\left(  \phi\right) + 6\frac{k}{a^{2}} \right)  \right),\label{eq:LagQ}
\end{equation}
and by varying Eq.~\eqref{eq:LagQ} with respect to $\{N, a, \phi, \Psi\}$, imposing $N=1$, we obtain the following cosmological field equations
\begin{subequations}
\begin{align}
    &\frac{1}{2} \omega_0\dot{\phi}^2+6ka^{-2}(1-\dot{\Psi}\dot{\phi})-\frac{6 \dot{\phi} a^{-2}}{\dot{\Psi}}+6 H^2+V(\phi)=0\label{eq:FRWQN},\\
    &12 \dot{H}+6ka^{-2}(1+\dot{\Psi}\dot{\phi})-\frac{6 \dot{\phi} a^{-2}}{\dot{\Psi}}+12 H \dot{\phi}+18 H^2+3 V(\phi)-\frac{3}{2} \omega_0\dot{\phi}^2=0\label{eq:FRWQa},\\
     &\omega_0\ddot{\phi}+\frac{1}{2} \omega_0\dot{\phi}^2+\dot{V}(\phi)+V(\phi)+6ka^{-2}(1-\ddot{\Psi}-H\dot{\Psi})+\frac{6 H a^{-2}}{\dot{\Psi}}-\frac{6 \ddot{\Psi} a^{-2}}{\dot{\Psi}^2}+3 \omega_0 H \dot{\phi}-6 H^2=0\label{eq:FRWQphi},\\
      &\frac{12 \ddot{\Psi} \dot{\phi}}{\dot{\Psi}^3}-\frac{6 \dot{\phi}^2}{\dot{\Psi}^2}-\frac{6 \ddot{\phi}}{\dot{\Psi}^2}-6 kH \dot{\phi}-\frac{6 H \dot{\phi}}{\dot{\Psi}^2}-6 k\ddot{\phi}-6 k\dot{\phi}^2=0\label{eq:FRWQpsi}.
\end{align}
\end{subequations}


\section{The autonomous systems}\label{sec3}

In this section, we derive the dynamical equations of the scalar field within the framework of the geometrical trinity of gravity, then focusing on $R$, $T$, and $Q$.  To investigate the linear stability of the resulting cosmological models, we subsequently recast the field equations into autonomous dynamical systems using

\begin{itemize}
    \item[-] dimensionless variables, \emph{ad hoc} formulated with the aim at recasting them into a first-order differential autonomous system;
    \item[-] the corresponding critical points, identified by setting each differential equation to zero.
\end{itemize}

To this end, we present the dynamical systems for each gravitational framework and their corresponding critical points.

\subsection{Analysis with the curvature}

Considering a general scalar field potential, the dimensionless variables for general relativity theory of gravity are defined as \cite{Kerachian:2021ses,Paliathanasis:2022pgu}
\begin{align}
&x = \frac{\dot{\phi}}{\sqrt{H^2 + \left|k\right|a^{-2}}}, \hspace{4mm} y = \frac{\hat{V}}{H^2 + \left|k\right|a^{-2}}, \\
&\eta = \frac{H}{\sqrt{H^2 + \left|k\right|a^{-2}}}, \hspace{4mm} \lambda = \frac{\hat{V}_{,\phi}}{\hat{V}},
\end{align}
where only the parameter $\eta$ is a priori confined in a finite range, i.e., $\eta \in \left[-1,1\right]$.

The advantages of using the above variables can be summarized below.

\begin{itemize}
    \item[-] In the presence of spatial curvature, normalization is performed using the factor $\sqrt{H^{2} + \left|k\right|a^{-2}}$, and Eqs. ~\eqref{eq:FRWRN}-\eqref{eq:FRWRphi} can be rewritten in terms of the variables defined above.
\item[-] They allow for open, close and flat universe, including the signs of $\dot a$, $\dot \phi$ and $\hat V$.
\item[-] The fourth variable is the derivative of a logarithm, therefore with the prescription of an exponential potential, $\lambda$ remains constant, significantly simplifying the overall description.
\item[-] The corresponding autonomous system appears extremely manageable and compact, depending, moreover, on the free parameters of the initial Lagrangian.
\end{itemize}

The autonomous system that describes the dynamics of the scalar field is given by the equations provided as follows
\begin{subequations}
\begin{align}
x' &= -\Bigg[ |k| \bigg(
24 \eta^2 + (\omega_0 - 2)\omega_0 \eta x^3 + 2x^2 (2\omega_0 \eta^2 + 5\omega_0 - 12) + 2\eta x \left( -2(\omega_0 - 6)\eta^2 + (2\lambda - \omega_0 + 2)y + 8\omega_0 - 12 \right) \nonumber \\
&\quad + (8\lambda - 4)y \bigg)+ 4k(\eta^2 - 1)\left( (\omega_0 - 6)\eta x - 6 \right) \Bigg]/\left(8 (\omega_0 - 3)|k|\right), \\
y' &= -y\Bigg[ |k| \bigg(
x (4 \omega_0 \eta^2 - 4\lambda (\omega_0 - 3)) + (\omega_0 - 2)\omega_0 \eta x^2 + 2\eta \left( -2(\omega_0 - 6)\eta^2 + (2\lambda - \omega_0 + 2)y - 4(\omega_0 - 3) \right) \bigg) \nonumber \\
&\quad + 4k(\omega_0 - 6)\eta(\eta^2 - 1) \Bigg]/\left(4 (\omega_0 - 3)|k|\right), \\
\eta' &= -(\eta^2 - 1) \left(
|k| \left( -4(\omega_0 - 6)\eta^2 + 4\omega_0 \eta x + (\omega_0 - 2)\omega_0 x^2 + (4\lambda - 2\omega_0 + 4)y \right) + 4k(\omega_0 - 6)(\eta^2 - 1) \right)/\left(8 (\omega_0 - 3)|k|\right),
\end{align}
\end{subequations}

where we defined a new parameter $\tau$ such that $d\tau=\sqrt{H^{2}+\left|k\right|a^{-2}}dt$, and prime is the derivative with respect to $\tau$. This new parameter describes the cosmic evolution, and it can be interpreted as the number of e-foldings modified by the presence of spatial curvature. It reduces to the standard expression $N = \ln a$ in the case of a spatially flat universe, i.e., when $k = 0$.

With this description, we can study the behavior of the scalar field by solving the dynamical system. In this way, we may start from early times, given by negative values of $\tau$, in order to observe the evolution of the universe up to late times, where $\tau=0$.

In addition, since the dimensionless variables satisfy the following constraint equation
\begin{equation}
y=6 \frac{k}{|k|}\left(\eta^2-1\right)-6 \eta^2-6 \eta x-\frac{1}{2} \omega_0 x^2,\label{eq:RCon}
\end{equation}

\begin{table*}[htb!]
\begin{center}
\setlength{\tabcolsep}{1.2em}
\renewcommand{\arraystretch}{1.8}
\begin{tabular}{|c|c|c|c|c|c|}
\hline
\hline
{\bf Critical point} & {$\bf x$} & {$\bf \eta$} & {\bf Existence} & {$\bf q$} \\
\hline\hline
\cline{1-5}
\hline\hline
   $P_{R,0}^{k=\pm 1}$ & $-\frac{2 (\lambda -1)}{\lambda -\omega_0+2}$ & $1$ & $\lambda-\omega_{0}+2\neq0$ & $\frac{(\lambda -2) \lambda +\omega_0-2}{\lambda -\omega_0+2}$ \\
$P_{R,1}^{k=\pm 1}$ & $\frac{2 (\lambda -1)}{\lambda -\omega_0+2}$ & $-1$ & $\lambda-\omega_{0}+2\neq0$  & $\frac{(\lambda -2) \lambda +\omega_0-2}{\lambda -\omega_0+2}$ \\
$P_{R,2}^{k=1}$ & $-\frac{2}{\sqrt{2 \lambda -\omega_0+2}}$ & $\frac{\lambda }{\sqrt{2 \lambda -\omega_0+2}}$ & $2 \lambda -\omega_0+2>0$ & $0$\\
$P_{R,3}^{k= 1}$ & $\frac{2}{\sqrt{2 \lambda -\omega_0+2}}$ & $-\frac{\lambda }{\sqrt{2 \lambda -\omega_0+2}}$ & $2 \lambda -\omega_0+2>0$ & $0$\\
$P_{R,2}^{k=-1}$ & $-\frac{2}{\sqrt{2 \lambda ^2-2 \lambda +\omega_0-2}}$ & $\frac{\lambda }{\sqrt{2 \lambda ^2-2 \lambda +\omega_0-2}}$ &  $2 \lambda ^2-2 \lambda +\omega_0-2>0$ & $0$\\
$P_{R,3}^{k=-1}$ & $\frac{2}{\sqrt{2 \lambda ^2-2 \lambda +\omega_0-2}}$ & $-\frac{\lambda }{\sqrt{2 \lambda ^2-2 \lambda +\omega_0-2}}$ & $2 \lambda ^2-2 \lambda +\omega_0-2>0$ & $0$\\
$P_{R,4}^{k=\pm 1}$ & $\frac{2 \left(-\sqrt{3} \sqrt{3-\omega_0}-3\right)}{\omega_0}$ & $1$ & $3-\omega_0\geq0\land \omega_0\neq 0$ & $\frac{2 \left(\omega_0-\sqrt{9-3 \omega_0}-3\right)}{\omega_0}$\\
$P_{R,5}^{k=\pm 1}$ & $\frac{2 \left(\sqrt{3} \sqrt{3-\omega_0}+3\right)}{\omega_0}$ & $-1$ & $3-\omega_0\geq0\land \omega_0\neq 0$ & $\frac{2 \left(\omega_0-\sqrt{9-3 \omega_0}-3\right)}{\omega_0}$\\
$P_{R,6}^{k=\pm 1}$ & $\frac{2 \left(\sqrt{3} \sqrt{3-\omega_0}-3\right)}{\omega_0}$ & $1$ & $3-\omega_0\geq0\land \omega_0\neq 0$ & $\frac{2 \left(\omega_0+\sqrt{9-3 \omega_0}-3\right)}{\omega_0}$\\
$P_{R,7}^{k=\pm 1}$ & $\frac{2 \left(3-\sqrt{3} \sqrt{3-\omega_0}\right)}{\omega_0}$ & $-1$ & $3-\omega_0\geq0\land \omega_0\neq 0$ & $\frac{2 \left(\omega_0+\sqrt{9-3 \omega_0}-3\right)}{\omega_0}$\\
$P_{R,8}^{k=-1}$ & $0$ & $\frac{1}{\sqrt{2}}$ & Always & $0$ \\
$P_{R,9}^{k=-1}$ & $0$ & $-\frac{1}{\sqrt{2}}$ & Always & $0$ \\
\hline\hline

\hline\hline

\end{tabular}
\caption{The critical points for the scalar field system in general relativity with exponential potential are given along with their existence conditions and deceleration parameter.}\label{tab:CritR}
\end{center}
\end{table*}

we can reduce the dimensionality of the system, from three to two first-order differential equations, i.e., $x'$ and $\eta'$. This constraint arises by substituting the dimensionless variables into Eq.~\eqref{eq:FRWRN}, yielding a consistent relation that simplifies the dynamical system.

We can also introduce the \emph{deceleration parameter} $q=-1-\frac{\dot{H}}{H^{2}}$ to determine whether the universe is undergoing an acceleration phase \cite{Capozziello:2021xjw}, i.e.,

\begin{equation}
    q=-\frac{|k|  \left(-4 (\omega_0-6) \eta^2+4 \omega_0\eta x+(\omega_0-2) \omega_0 x^2+(4 \lambda -2 \omega_0+4) y\right)+4 k(\omega_0-6) \left(\eta^2-1\right)}{8 (\omega_0-3) |k|  \eta^2}.
\end{equation}

The critical points of the system are obtained by setting $x' = \eta' = 0$ for the cases $k=1$ and $k=-1$, which correspond to a closed and an open universe, respectively. These points characterize equilibrium states in cosmic evolution, where the phase space variables remain constant over time. Then, we compute the corresponding deceleration parameter, which provides information about the acceleration or deceleration of universe's expansion history. The results for each scenario are summarized in Tab.~\ref{tab:CritR}.

Furthermore, we identify critical points where $\eta = \pm 1$, which correspond to a regime in which the effects of spatial curvature become negligible, and this behavior is particularly relevant in the late time evolution of the universe, or in the verly early stages of the universe near to the cosmological singularity.

For a positive value of the spatial curvature, that is, $k =1$, the field equations admit eight stationary points, which can describe an expanding solution if $\eta > 0$ and a collapsing one if $\eta < 0$.

The points $P_{R,2}^{k=1}$, $P_{R,3}^{k=1}$ describe solutions where the curvature term and the scalar field are affected by the values of $\omega_{0}$ and $\lambda$. In particular, the rescaled dimensionless Hubble variable can satisfy $\eta \neq \pm 1$, corresponding to a non-flat universe that asymptotically approaches a state with a vanishing deceleration parameter.

In contrast, the rest of the stationary points indicate asymptotically spatially flat universes. On the one hand, for the points $P_{R, 0}^{k=1}$ and $P_{R, 1}^{k=1}$, we calculate the deceleration parameter as $q=\frac{(\lambda -2) \lambda +\omega_0-2}{\lambda -\omega_0+2}$. On the other hand, for the points $P_{R,4}^{k=1}$, $P_{R,5}^{k=1}$, $P_{R,6}^{k=1}$ and $P_{R,7}^{k=1}$, the deceleration parameter is provided by $q=\frac{2 \left(\omega_0+\sqrt{9-3 \omega_0}-3\right)}{\omega_0}$.

In the case of negative spatial curvature, i.e., when imposing $k = -1$, the dynamical system exhibits ten critical points, two more than in the positively curved scenario. The points $P_{R,2}^{k=-1}$ and $P_{R,3}^{k=-1}$ differ from their counterparts in the $k =1$ case, although they share the same deceleration parameter. The additional points, $P_{R,8}^{k=-1}$ and $P_{R,9}^{k=-1}$, are characterized by a vanishing dimensionless rescaled field velocity $x$, while the rescaled Hubble parameter satisfies $\eta \neq \pm 1$. This behavior indicates a residual curvature contribution that remains relevant at the end of the scalar field evolution. In these cases, the deceleration parameter is zero, as a consequence of the non-negligible spatial curvature.

At this stage, a degeneracy can be identified between closed and open universes for the subset of points associated with asymptotically flat dynamics, i.e, $\eta=\pm1$. These critical points share identical values of $x$ and the deceleration parameter, identifying the same cosmological scenario. However, they are excluded as viable critical points for a dark energy candidate since they exhibit a vanishing deceleration parameter, denoting the absence of accelerated expansion.

\subsection{Analysis with torsion}

In the case of teleparallel theory of gravity, the dimensionless variables are the same as those used in general relativity background. However, the results appear different. That is due to the difference between the boundary term which relates the Ricci scalar and the torsion scalar.

Indeed, considering the torsion scalar instead of Ricci one, we get the dynamical system composed by the following set of equations
\begin{subequations}
\begin{align}
        &x'=-\frac{| k|  \left(-48 \eta^2+\omega_0^2 \eta x^3+x^2 \left(4 \omega_0-8 \omega_0 \eta^2\right)-2 \omega_0 \eta x \left(2 \eta^2+y-8\right)+8 (\lambda +1) y\right)+4 k \left(\eta^2-1\right) (\omega_0 \eta x-12)}{8 \omega_0 | k| },\\
        &y'=\frac{1}{4} y\left(2 \eta\left(-\frac{2 k \left(\eta^2-1\right)}{|k|}+2 \eta^2+y+4\right)+4 x \left(\lambda +2 \eta^2\right)-\omega_0 \eta x^2\right),\\
        &\eta'=\frac{\left(\eta^2-1\right) \left(| k|\left(4 \eta^2+8 \eta x-\omega_0 x^2+2 y\right)-4 k \left(\eta^2-1\right)\right)}{8|k|},
\end{align}
\end{subequations}
where prime indicates again the derivative with respect to $\tau$.

The above system can be reduced to a system of only two differential equations by rewriting the constraint in Eq.~\eqref{eq:FRWTN} in terms of our dimensionless variables, i.e.,
\begin{equation}
    y=6 \frac{k}{|k|}\left(\eta^2-1\right)-6 \eta^2-\frac{1}{2} \omega_0 x^2,
\end{equation}
and the deceleration parameter is computed as
\begin{equation}
    q=\frac{|k|  \left(4 \eta^2+8 \eta x-\omega_0 x^2+2 y\right)-4 k \left(\eta^2-1\right)}{8 |k|\eta^2}.
\end{equation}

\begin{table*}[htb!]
\begin{center}
\setlength{\tabcolsep}{1.2em}
\renewcommand{\arraystretch}{1.8}
\begin{tabular}{|c|c|c|c|c|c|}
\hline
\hline
{\bf Critical point} & {$\bf x$} & {$\bf \eta$} & {\bf Existence} & {$\bf q$} \\
\hline\hline
\cline{1-5}
\hline\hline
$P_{T,0}^{k=\pm1}$ & $\frac{2 (\lambda +2)}{\omega_0}$ & $1$ & $\omega_{0}\neq 0$ & $-\frac{\lambda  (\lambda +2)+\omega_0}{\omega_0}$\\
$P_{T,1}^{k=\pm1}$ & $-\frac{2 (\lambda +2)}{\omega_0}$ & $-1$ &  $\omega_{0}\neq 0$ & $-\frac{\lambda  (\lambda +2)+\omega_0}{\omega_0}$\\
$P_{T,2}^{k=1}$ & $-\frac{2}{\sqrt{-2 \lambda -\omega_0}}$ & $\frac{\lambda }{\sqrt{-2 \lambda -\omega_0}}$ & $-2\lambda-\omega_0>0$  & $0$ \\
$P_{T,3}^{k=1}$ & $\frac{2}{\sqrt{-2 \lambda -\omega_0}}$ & $-\frac{\lambda }{\sqrt{-2 \lambda -\omega_0}}$ & $-2\lambda-\omega_0>0$ & $0$ \\
$P_{T,2}^{k=-1}$ & $-\frac{2}{\sqrt{2 \lambda  (\lambda +1)+\omega_0}}$ & $\frac{\lambda }{\sqrt{2 \lambda  (\lambda +1)+\omega_0}}$ & $2 \lambda  (\lambda +1)+\omega_0>0$ & $0$ \\
$P_{T,3}^{k=-1}$ & $\frac{2}{\sqrt{2 \lambda  (\lambda +1)+\omega_0}}$ & $-\frac{\lambda }{\sqrt{2 \lambda  (\lambda +1)+\omega_0}}$ & $2 \lambda  (\lambda +1)+\omega_0>0$ & $0$ \\
$P_{T,4}^{k=\pm1}$ & $2\sqrt{\frac{3}{-\omega_0}}$ & $-1$ & $\omega_0<0$ & $2-2\sqrt{\frac{3}{-\omega_0}}$\\
$P_{T,5}^{k=\pm1}$ & $-2\sqrt{\frac{3}{-\omega_0}}$ & $1$ & $\omega_0<0$ & $2-2\sqrt{\frac{3}{-\omega_0}}$\\
$P_{T,6}^{k=\pm1}$ & $-2\sqrt{\frac{3}{-\omega_0}}$ & $-1$ & $\omega_0<0$ & $2+2\sqrt{\frac{3}{-\omega_0}}$\\
$P_{T,7}^{k=\pm1}$ & $2\sqrt{\frac{3}{-\omega_0}}$ & $1$ & $\omega_0<0$ & $2+2\sqrt{\frac{3}{-\omega_0}}$\\
$P_{T,8}^{k=1}$ & $2\sqrt{\frac{3}{-\omega_0}}$ & $\sqrt{\frac{-\omega_0}{3}}$ & $\omega_0<0$ & $0$\\
$P_{T,9}^{k=1}$ & $-2\sqrt{\frac{3}{-\omega_0}}$ &$-\sqrt{\frac{-\omega_0}{3}}$ & $\omega_0<0$ & $0$\\
$P_{T,8}^{k=-1}$ & $-\frac{6}{\sqrt{\omega_0 (2 \omega_0+3)}}$ & $-\frac{\omega_0}{\sqrt{\omega_0 (2 \omega_0+3)}}$ & $\omega_0 (2 \omega_0+3)>0$ & $0$\\

$P_{T,9}^{k=-1}$ & $\frac{6}{\sqrt{\omega_0 (2 \omega_0+3)}}$ & $\frac{\omega_0}{\sqrt{\omega_0 (2 \omega_0+3)}}$ & $\omega_0 (2 \omega_0+3)>0$ & $0$\\
\hline\hline

\end{tabular}
\caption{The critical points for the scalar field system in teleparallel gravity with exponential potential are given along with their existence conditions and deceleration parameter.}\label{tab:CritT}
\end{center}
\end{table*}

Here, as in general relativity, critical points are detected by imposing $x'=\eta'=0$, and they are shown in Tab. \ref{tab:CritT} along with the value of the deceleration parameter.

Specifically, we identify ten critical points for both closed and open universes. The points $P^{k=\pm1}_{T,0}$, $P^{k=\pm1}_{T,1}$, $P^{k=\pm1}_{T,4}$, $P^{k=\pm1}_{T,5}$, $P^{k=\pm1}_{T,6}$, and $P^{k=\pm1}_{T,7}$ indicate that the universe is flat in the final stages of its evolution, i.e., $\eta = \pm 1$. These points are the same for both $k =1$ and $k = -1$, meaning that, also within the framework of teleparallel gravity, there exists a degeneracy when a closed or open universe is assumed at the beginning of the dynamics. This degeneracy arises because the contribution from spatial curvature becomes negligible at late times, leading to equivalent asymptotic behavior in both scenarios. In particular, as we observed in the general relativity background, the critical points corresponding to $\eta = \pm 1$ are the same, independently of whether the analysis begins by imposing a closed or an open universe.

Regarding the points $P^{k=\pm1}_{T,2}$, $P^{k=\pm1}_{T,3}$, $P^{k=\pm1}_{T,8}$ and $P^{k=\pm1}_{T,9}$, the contribution from spatial curvature may become non-negligible toward the end of the dynamical evolution. These critical points attain different values depending on whether the analysis begins with a closed or open universe; however, the deceleration parameter remains the same in all cases and vanishes at these points. This implies that critical points where curvature effects persist are incompatible with scenarios in which the scalar field drives cosmic acceleration at late times. In contrast, at the points $P^{k=\pm 1}_{T,0}$ and $P^{k=\pm 1}_{T,1}$, the deceleration parameter is given by
$q = -\dfrac{\lambda(\lambda + 2) + \omega_0}{\omega_0}$.
Finally, for the points $P_{T,4}$ and $P_{T,5}$, it is given by
$q = 2-2\sqrt{\frac{3}{-\omega_0}}$,
while for $P_{T,6}$ and $P_{T,7}$, it takes the form
$q = 2+2\sqrt{\frac{3}{-\omega_0}}$. Moreover, we observe that, in the context of teleparallel gravity, condition $\omega_0 < 0$ holds for the points $P^{k=\pm 1}_{T,4}$, $P^{k=\pm 1}_{T,5}$, $P^{k=\pm 1}_{T,6}$, $P^{k=\pm 1}_{T,7}$, $P_{T,8}^{k=1}$, and $P_{T,9}^{k=1}$. This leads us to argue that the scalar field behaves as a phantom-like field, due to the negative kinetic
energy implied by the factor $\omega_0$.

\subsection{Analysis with non-metricity}

The dilaton model in the symmetric-teleparallel theory assumes a different configuration than the previous ones. Indeed, the use of the non-metricity scalar in the non-coincident gauge requires the introduction of an additional scalar field, as we deduce from Eq.~\eqref{eq:Qscalar}.

Therefore, we introduce a new set of dimensionless variables defined as
\begin{align}
&x = \frac{\dot{\phi}}{2\sqrt{3}H},\hspace{4mm}z =2\sqrt{3}\dot{\Psi}a^{2}H, \hspace{4mm} y = \frac{\sqrt{\hat{V}}}{\sqrt{6}H}, \\
&\Omega_k = \frac{k}{a^{2}H^{2}},\hspace{4mm} \lambda = \frac{\hat{V}_{,\phi}}{\hat{V}},
\end{align}
where $\lambda = \text{const.}$, as we consider again only an exponential potential.

Here, due to the non-linearity of the field equations, we employ the Hubble normalization, and the spatial curvature is treated as a dynamical variable to reduce the system's complexity.

Thus, having two new dimensionless variables concerning the additional scalar field $\Psi$ and the spatial curvature of the universe $k$, the dynamical system is provided by the following differential equations

\begin{subequations}
\begin{align}
&x'= \frac{x}{2} \left( \frac{96 \left( \sqrt{3} \left( -\omega_0 x^2 + (\lambda + 1)y^2 + \Omega_k - 1 \right) - \Omega_k z^2 + 12 \right)}{\left( \Omega_k z^2 + 12 \right)^2 - 48 \omega_0 x z} + 3 \left( -\omega_0 x^2 - \frac{4x}{z} + y^2 - 1 \right) + \Omega_k (3xz + 1) \right), \\
&y'= \frac{y \left( z \left( 2 \sqrt{3} (\lambda + 2) x - 3 \omega_0 x^2 + 3 y^2 + \Omega_k + 3 \right) + 3 x \Omega_k z^2 - 12x \right)}{2z}, \\
&\Omega_k'= \Omega_k \left( -3 \omega_0 x^2 + x \left( 3 \Omega_k z - \frac{12}{z} + 4 \sqrt{3} \right) + 3 y^2 + \Omega_k + 1 \right), \\
&z'= \Bigg( x \Big( z \Big( z \Big( -144 \omega_0 y^2 + \Omega_k \Big( z \Big( \Omega_k z \left( 3 \Omega_k z^2 + 4 \sqrt{3} z + 60 \right) + 96 \sqrt{3} \Big) -48 (\omega_0 - 3) \Big) - 48 \omega_0 \Big) + 576 \sqrt{3} \Big) - 1728 \Big) \nonumber \\
&\quad + 144 \omega_0^2 x^3 z^2 + \omega_0 x^2 z \left( 144 - z \left( \Omega_k z \left( 3 \Omega_k z^2 - 4 \sqrt{3} z + 216 \right) + 144 \sqrt{3} \right) \right)+ z \left( \Omega_k z^2 + 12 \right)\times \nonumber \\
&\quad  \times\left( 12 \left( 3 y^2 + \Omega_k - 3 \right) - 4 \sqrt{3} z \left( (\lambda + 1) y^2 + \Omega_k - 1 \right) + \Omega_k z^2 \left( 3 y^2 + \Omega_k + 5 \right) \right) \Bigg) \Big/ \left( 96 \omega_0 x z - 2 \left( \Omega_k z^2 + 12 \right)^2 \right),
\end{align}
\end{subequations}
where now prime differs from the previous stability analyses, since it stands for the derivative with respect to the number of e-foldings $N=\ln a$.

In this framework, the deceleration parameter is given by
\begin{equation}
    q = \frac{1}{2} \left(-3 \omega_0 x^2 + x \left(3 \Omega_k z - \frac{12}{z} + 4 \sqrt{3}\right) + 3 y^2 + \Omega_k + 1\right),
\end{equation}
and, analogous to $R$ and $T$ backgrounds, the system in $Q$ can be reduced to a three-dimensional dynamical system by imposing the following constraint
\begin{equation}
    y = \sqrt{\frac{-\omega_0 x^2 z + x\left(\Omega_k z^2 + 12\right) - (\Omega_k + 1) z}{z}},
\end{equation}
derived by replacing the new dimensionless variables in Eq.~\eqref{eq:FRWQN}.

At this stage, the critical points, presented in Tab.~\ref{tab:CritQ}, are obtained simultaneously by solving $x' = \Omega_{k}' = z' = 0$. These points are determined by incorporating the above constraint to reduce the dimensionality of the system. This reduction leads to a degeneracy between points $P_{Q,2}$ and $P_{Q,3}$, as they differ only in the sign of the $y$ value, which is not considered here.

We also observe that, in the symmetric-teleparallel scenario, the critical points that allow for a non-flat FRW metric at the end of the dynamical evolution—namely, $P_{Q,4}$, $P_{Q,5}$, $P_{Q,6}$, and $P_{Q,7}$—are characterized again by a vanishing deceleration parameter. Consequently, these points are not physically acceptable if we aim to interpret the scalar field as a candidate for dark energy.

The only critical points that allow for an accelerated expansion of the universe, under suitable choices of the parameters $\lambda$ and $\omega_0$, are $P_{Q,0}$, $P_{Q,1}$, $P_{Q,2}$, $P_{Q,3}$, and $P_{Q,4}$.
For the point $P_{Q,0}$, the deceleration parameter is given by
$q = \dfrac{2(\omega_0 + 3)}{3(\omega_0 + \sqrt{1 - \omega_0} + 1)}$,
while for $P_{Q,1}$, it is given by
$q = \dfrac{2(\omega_0 + \sqrt{1 - \omega_0} + 1)}{3\omega_0}$.
For the points $P_{Q,3}$ and $P_{Q,4}$, the deceleration parameter is identical and takes the form
$q = \dfrac{2(\lambda - 1)}{3\lambda + 2}$.
At all of these points, the dimensionless variable associated with spatial curvature vanishes, indicating that the universe is asymptotically flat.

\begin{table*}[htb!]
\begin{center}
\setlength{\tabcolsep}{0.005em}
\renewcommand{\arraystretch}{1.8}
\begin{tabular}{|c|c|c|c|c|c|}
\hline
\hline
{\bf Critical point} & {$\bf x$} & {$\bf z$} & {$\bf \Omega_{k}$} & {\bf Existence} & {$\bf q$} \\
\hline\hline
\cline{1-6}
\hline\hline
$P_{Q,0}$ & $\frac{1}{\sqrt{3} \left(\sqrt{1-\omega_0}+1\right)}$ & $-\frac{6 \sqrt{3} \left(\sqrt{1-\omega_0}-2\right)}{\omega_0+3}$ & $0$ &  $\omega_0\leq1\land \omega_0\neq -3$ & $\frac{2 (\omega_0+3)}{3 \left(\omega_0+\sqrt{1-\omega_0}+1\right)}$\\
$P_{Q,1}$ & $\frac{\sqrt{1-\omega_0}+1}{\sqrt{3} \omega_0}$ & $\frac{6 \sqrt{3} \left(\sqrt{1-\omega_0}+2\right)}{\omega_0+3}$ & $0$ & $\omega_0\leq1\land \omega_0\neq \{0,-3\}$  & $\frac{2 \left(\omega_0+\sqrt{1-\omega_0}+1\right)}{3 \omega_0}$\\
$P_{Q,2}$ & $-\frac{5}{\sqrt{3} (3 \lambda +2)}$ & $-\frac{4 \sqrt{3} (3 \lambda +2)}{\lambda  (3 \lambda +8)+5 \omega_0+4}$ & $0$ & $\lambda\neq -\frac{2}{3}\land \lambda(3\lambda+8)+5\omega_0+4\neq0$  & $\frac{2 (\lambda -1)}{3 \lambda +2}$ \\
$P_{Q,3}$ & $ -\frac{5}{\sqrt{3} (3 \lambda +2)}$& $-\frac{4 \sqrt{3} (3 \lambda +2)}{\lambda  (3 \lambda +8)+5 \omega_0+4}$ & $0$ & $\lambda\neq -\frac{2}{3}\land \lambda(3\lambda+8)+5\omega_0+4\neq0$ & $\frac{2 (\lambda -1)}{3 \lambda +2}$ \\
$P_{Q,4}$ & $-\frac{1}{\sqrt{3} \lambda }$ & $-\frac{2 \sqrt{3} \lambda  \left(\sqrt{\Delta}-2\right)}{\lambda  (\lambda +2)+\omega_0}$ & $-\frac{4 \sqrt{\Delta}+\lambda  (\lambda +2)+\omega_0+8}{\lambda ^2} $ & $\Delta\geq0\land\lambda\neq0 \land \lambda(\lambda+2)+\omega_0\neq0$ & $0$\\

$P_{Q,5}$ & $-\frac{1}{\sqrt{3} \lambda } $ &$\frac{2 \sqrt{3} \lambda  \left(\sqrt{\Delta}+2\right)}{\lambda  (\lambda +2)+\omega_0}$ & $-\frac{-4 \sqrt{\Delta}+\lambda  (\lambda +2)+\omega_0+8}{\lambda ^2}$ & $\Delta\geq0\land\lambda\neq0 \land \lambda(\lambda+2)+\omega_0\neq0$ & $0$\\

$P_{Q,6}$ & $\frac{\sqrt{3} \left(\omega_0+2 \sqrt{\omega_{0}^{2}-9\omega_{0}}\right)}{(\omega_0-12) \omega_0}$ &$-\frac{2 \sqrt{3} \left(\sqrt{\omega_{0}^{2}-9\omega_{0}}-6\right)}{\omega_0+3}$ & $ -\frac{\omega_0^2-9 \omega_0+12 \sqrt{\omega_{0}^{2}-9\omega_{0}}+36}{(\omega_0-12)^2}$ &  $\omega_0\leq0\lor \omega_0\geq9 \land \omega_0 \neq \{-3,12\}$ & $0$\\

$P_{Q,7}$ & $\frac{\sqrt{3} \left(\omega_0-2 \sqrt{\omega_{0}^{2}-9\omega_{0}}\right)}{(\omega_0-12) \omega_0}$ & $\frac{2 \sqrt{3} \left(\sqrt{\omega_{0}^{2}-9\omega_{0}}+6\right)}{\omega_0+3}$ & $\frac{-\omega_0^2+9 \omega_0+12 \sqrt{\omega_{0}^{2}-9\omega_{0}}-36}{(\omega_0-12)^2}$ & $\omega_0\leq0\lor \omega_0\geq9 \land \omega_0 \neq \{-3,12\}$ & $0$\\
\hline\hline

\end{tabular}
\caption{The critical points for the scalar field system in non-metricity gravity with exponential potential are given along with their existence conditions and deceleration parameter, where $\Delta=\lambda  (\lambda +2)+\omega_0+4$.}\label{tab:CritQ}
\end{center}
\end{table*}


\section{The stability of the systems}\label{sec4}

In this section, we perform a linear stability analysis of our scalar field model within the framework of the trinity of gravity. Specifically, we examine the stability of critical points in our dynamical systems, with particular focus on attractor points—those critical points where the dynamics ultimately converge. Critical points are classified as follows \cite{Copeland:2006wr,Bahamonde:2017ize}:
\begin{itemize}
    \item[-] \emph{\textbf{Stable point}}: All the real parts of the eigenvalues of the Jacobian matrix are negative.
    \item[-] \emph{\textbf{Unstable point}}: All the real parts of the eigenvalues of the Jacobian matrix are positive.
    \item[-] \emph{\textbf{Saddle point}}: At least one real part of an eigenvalue of the Jacobian matrix is positive.
\end{itemize}
A critical point is considered an attractor for the system only if it is a stable node or a stable spiral.

The characteristics of these points, for each gravity scenario in both closed and open universes, are presented below.

\subsection{Analysis with curvature}

To study the dynamics of the autonomous systems in general relativity background, we analyze the small perturbations around critical points, i.e.,
\begin{eqnarray}
\left(
\begin{array}{c}
\delta x' \\
\delta \eta' \\
\end{array}
\right) = {\mathcal J} \left(
\begin{array}{c}
\delta x \\
\delta \eta \\
\end{array}
\right) \,,
\label{eq:pert0}
\end{eqnarray}
where $\mathcal{J}$ is the Jacobian matrix. This matrix is computed at the critical points, and it is provided by
\begin{eqnarray}
\mathcal{J}=\left( \begin{array}{cc}
\frac{\partial x'}{\partial x}& \frac{\partial x'}{\partial \eta}\\
\frac{\partial \eta'}{\partial x}& \frac{\partial \eta'}{\partial \eta}\\
\end{array} \right)_{P_{c}}\,,
\label{eq:J0}
\end{eqnarray}
where $P_{c}=\left(x_{c},\eta_{c}\right)$ denotes the critical point, indicated by the subscript $c$.

For both closed and open universes, the eigenvalues corresponding to each critical point are listed in Tab.~\ref{tab:EigenR}. In particular, we focus on the attractor points of the autonomous system,
\begin{equation} \left(x(t), \eta(t)\right) \rightarrow \left(x_c, \eta_c\right) \quad \text{as} \quad t \rightarrow \infty, \end{equation}
representing the critical points toward which the system evolves asymptotically.

\begin{table*}[htb!]
\begin{center}
\setlength{\tabcolsep}{1.2em}
\renewcommand{\arraystretch}{1.8}
\begin{tabular}{|c|c|c|}
\hline
\hline
{\bf Critical point} & {\bf Eigenvalues} & {\bf Stability} \\
\hline\hline
\cline{1-3}
\hline\hline
   $P_{R,0}^{k=\pm 1}$ & $\left\{\frac{2 ((\lambda -2) \lambda +\omega_0-2)}{\lambda -\omega_0+2},\frac{(\lambda -2) \lambda +3 \omega_0-8}{\lambda -\omega_0+2}\right\}$ & $-$ \\
$P_{R,1}^{k=\pm 1}$ & $\left\{-\frac{2 ((\lambda -2) \lambda +\omega_0-2)}{\lambda -\omega_0+2},-\frac{(\lambda -2) \lambda +3 \omega_0-8}{\lambda -\omega_0+2}\right\}$ & $-$ \\
$P_{R,2}^{k=1}$ & $\left\{\frac{i \sqrt{3 (\lambda -2) \lambda +4 \omega_0-9}-\lambda +1}{\sqrt{2 \lambda -\omega_0+2}},\frac{-i \sqrt{3 (\lambda -2) \lambda +4 \omega_0-9}-\lambda +1}{\sqrt{2 \lambda -\omega_0+2}}\right\}$ & Saddle if
$\omega_0<-\lambda ^2+2 \lambda +2$\\
$P_{R,3}^{k=1}$ & $\left\{\frac{i \sqrt{3 (\lambda -2) \lambda +4 \omega_0-9}+\lambda -1}{\sqrt{2 \lambda -\omega_0+2}},\frac{-i \sqrt{3 (\lambda -2) \lambda +4 \omega_0-9}+\lambda -1}{\sqrt{2 \lambda -\omega_0+2}}\right\}$ & Saddle if $\omega_0<-\lambda ^2+2 \lambda +2$ \\
$P_{R,2}^{k=-1}$ & $\left\{\frac{-i \sqrt{3 (\lambda -2) \lambda +4 \omega_0-9}-\lambda +1}{\sqrt{2 (\lambda -1) \lambda +\omega_0-2}},\frac{i \sqrt{3 (\lambda -2) \lambda +4 \omega_0-9}-\lambda +1}{\sqrt{2 (\lambda -1) \lambda +\omega_0-2}}\right\}$ & $-$ \\
$P_{R,3}^{k=-1}$ & $\left\{\frac{-i \sqrt{3 (\lambda -2) \lambda +4 \omega_0-9}+\lambda -1}{\sqrt{2 (\lambda -1) \lambda +\omega_0-2}},\frac{i \sqrt{3 (\lambda -2) \lambda +4 \omega_0-9}+\lambda -1}{\sqrt{2 (\lambda -1) \lambda +\omega_0-2}}\right\}$ & $-$\\
$P_{R,4}^{k=\pm1}$ & $\left\{\frac{4 \left(\omega_0-\sqrt{9-3 \omega_0}-3\right)}{\omega_0},-\frac{2 \left(\lambda  \left(\sqrt{9-3 \omega_0}+3\right)-3 \omega_0+2 \sqrt{9-3 \omega_0}+6\right)}{\omega_0}\right\}$ & $-$ \\
$P_{R,5}^{k=\pm1}$ & $\left\{-\frac{4 \left(\omega_0+\sqrt{9-3 \omega_0}-3\right)}{\omega_0},-\frac{2 \left(\lambda  \left(\sqrt{9-3 \omega_0}-3\right)+3 \omega_0+2 \sqrt{9-3 \omega_0}-6\right)}{\omega_0}\right\}$ & $-$\\
$P_{R,6}^{k=\pm1}$ & $\left\{\frac{4 \left(\omega_0+\sqrt{9-3 \omega_0}-3\right)}{\omega_0},\frac{2 \left(\lambda  \left(\sqrt{9-3 \omega_0}-3\right)+3 \omega_0+2 \sqrt{9-3 \omega_0}-6\right)}{\omega_0}\right\}$ & $-$\\
$P_{R,7}^{k=\pm1}$ & $\left\{\frac{4 \left(-\omega_0+\sqrt{9-3 \omega_0}+3\right)}{\omega_0},\frac{2 \left(\lambda  \left(\sqrt{9-3 \omega_0}+3\right)-3 \omega_0+2 \sqrt{9-3 \omega_0}+6\right)}{\omega_0}\right\}$ & $-$\\
$P_{R,8}^{k=-1}$ & $\left\{-\sqrt{2},\sqrt{2}\right\}$ & Saddle \\
$P_{R,9}^{k=-1}$ & $\left\{-\sqrt{2},\sqrt{2}\right\}$ & Saddle \\
\hline\hline

\hline\hline

\end{tabular}
\caption{The eigenvalues for the scalar field system in general relativity background with exponential potential are given along with their stability conditions. In cases where the analytical stability analysis is not provided, the results are illustrated in Fig.~\ref{fig:stabcond1}.}\label{tab:EigenR}
\end{center}
\end{table*}

The key differences between closed and open universes arise from four critical points. Specifically, for $k=-1$, two additional saddle points appear, namely $P_{R,8}^{k=-1}$ and $P_{R,9}^{k=-1}$. Furthermore, the stability analysis of the points $P_{R,2}^{k=-1}$ and $P_{R,3}^{k=-1}$ differs due to variations in their eigenvalues. All other critical points remain unchanged, indicating a degeneracy between the closed and open universe cases from a stability perspective.

The stability of these critical points is analyzed by selecting appropriate values of $\lambda$ and $\omega_0$. In particular, for $\omega_{0} > 0$, the scalar field behaves as a quintessence-like field, while for $\omega_{0} < 0$, it corresponds to a phantom-like field. To illustrate this, we identify the regions in the $\{\lambda, \omega_0\}$ parameter space that characterize the stability conditions of the critical points, as shown in Fig.~\ref{fig:stabcond1}.

\begin{figure*}[htb!]
\centering
\subfigure[Phase-space portrait graph for $\omega_{0}=-8$ and $\lambda=2$.\label{1PSRk1}]
{\includegraphics[height=0.3\hsize,clip]{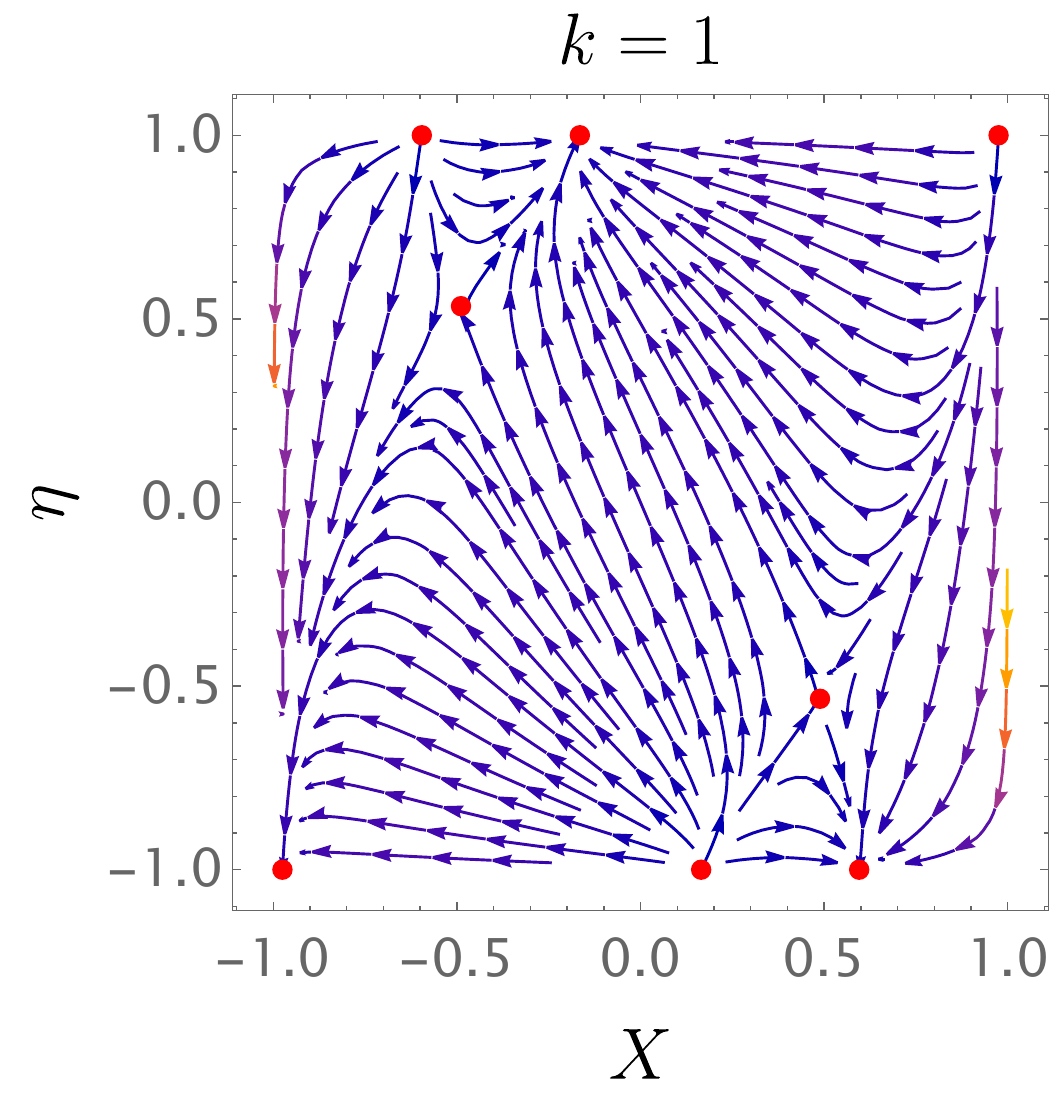}}
\subfigure[Phase-space portrait graph for $\omega_{0}=-6$ and $\lambda=1$.\label{2PSRk1}]
{\includegraphics[height=0.3\hsize,clip]{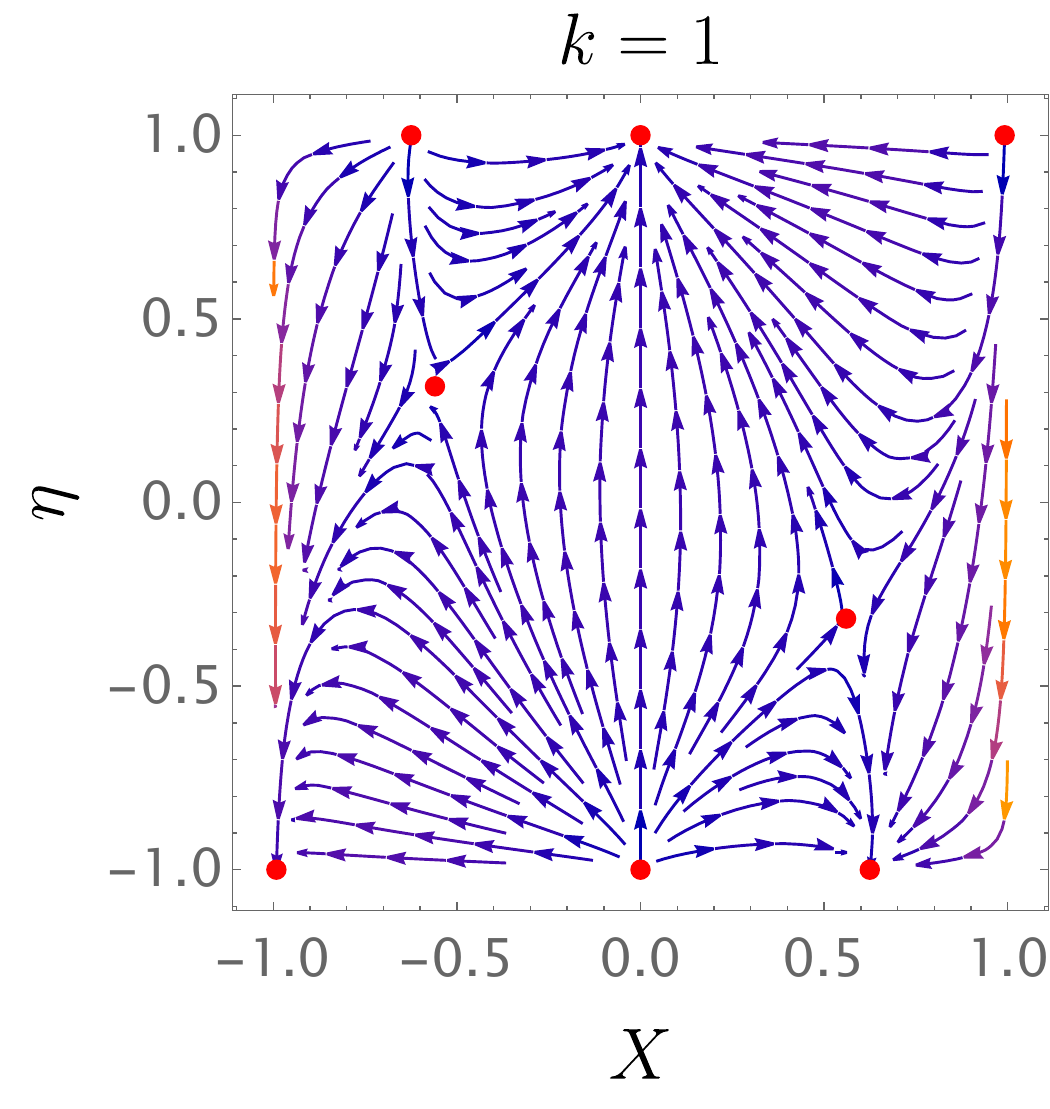}}
\subfigure[Phase-space portrait graph for $\omega_{0}=-4$ and $\lambda=0$.\label{3PSRk1}]
{\includegraphics[height=0.3\hsize,clip]{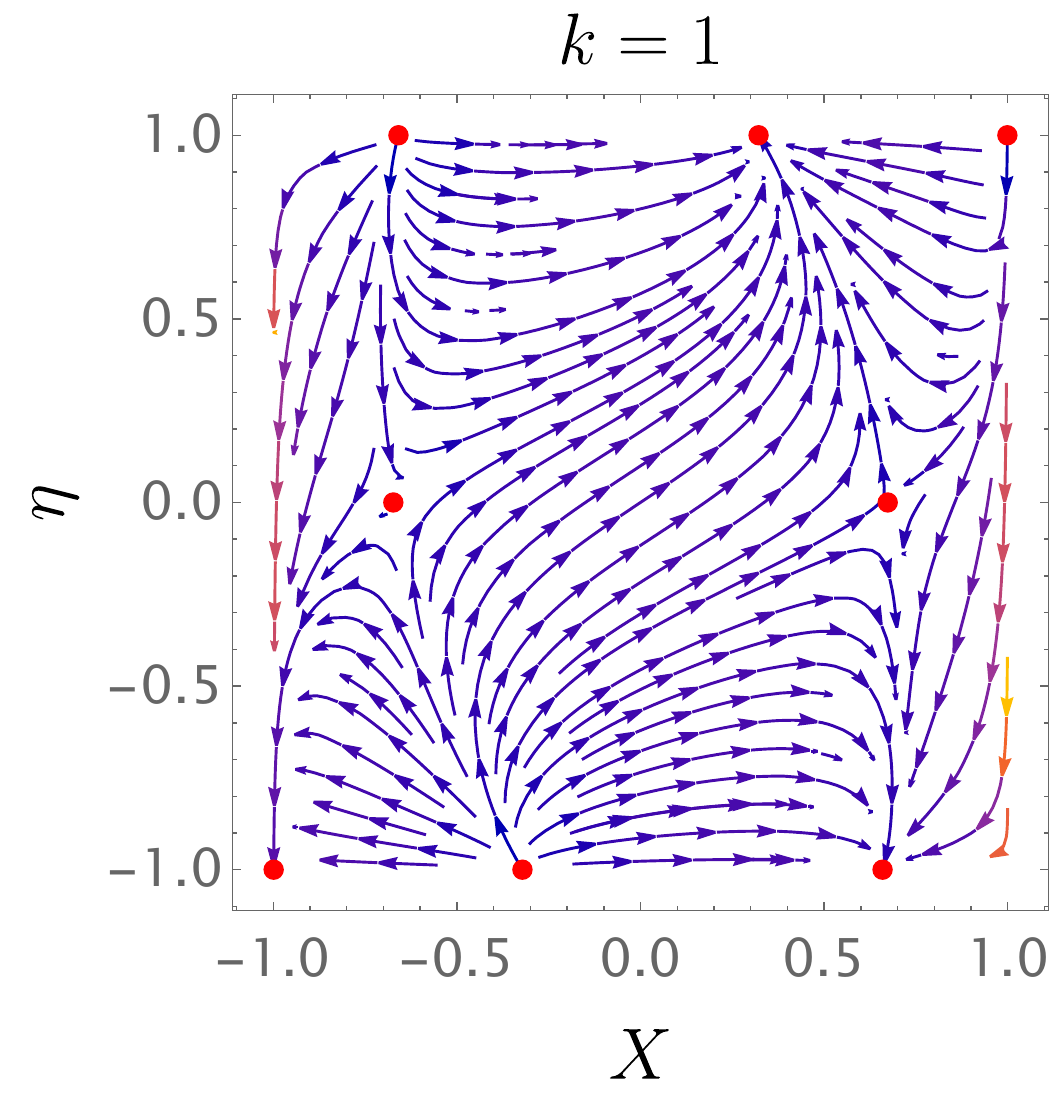}}
\subfigure[Phase-space portrait graph for $\omega_{0}=8$ and $\lambda=2$.\label{1PSRk-1}]
{\includegraphics[height=0.3\hsize,clip]{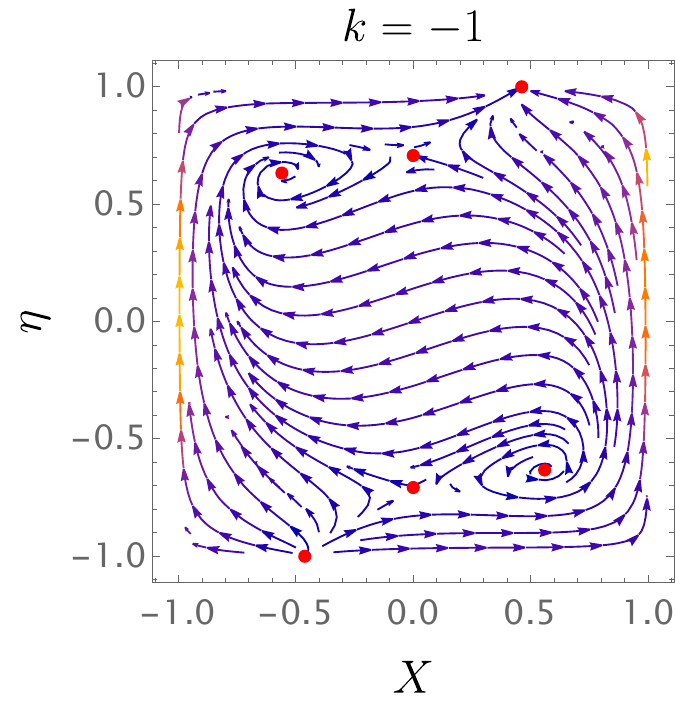}}
\subfigure[Phase-space portrait graph for $\omega_{0}=-6$ and $\lambda=1$.\label{2PSRk-1}]
{\includegraphics[height=0.3\hsize,clip]{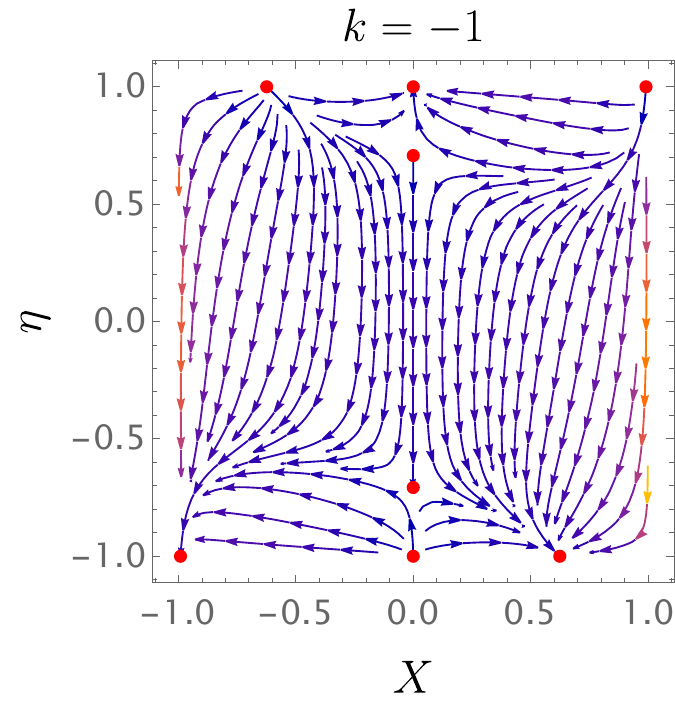}}
\subfigure[Phase-space portrait graph for $\omega_{0}=-4$ and $\lambda=0$.\label{3PSRk-1}]
{\includegraphics[height=0.3\hsize,clip]{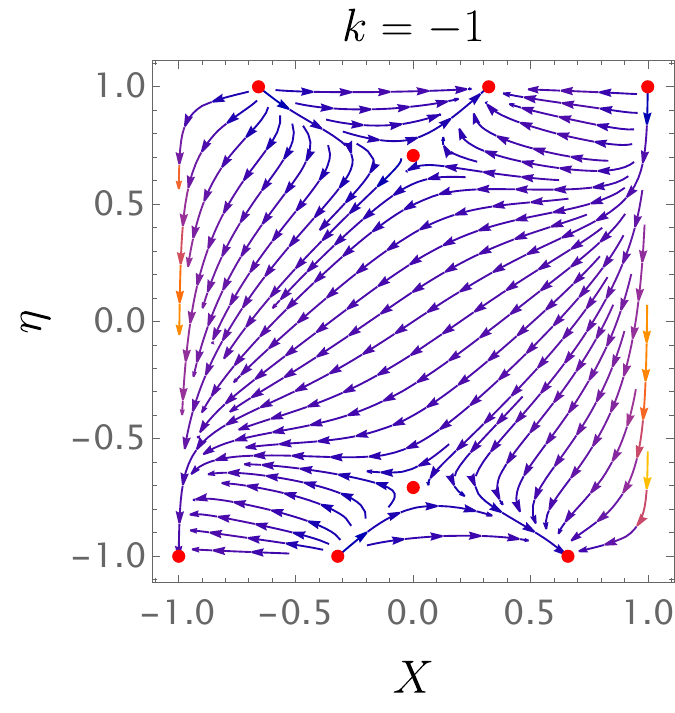}}
\caption{Phase-space portrait graphs for scalar field with the deceleration parameter in general relativity framework.}
\label{fig:PSR}
\end{figure*}

\begin{figure*}[htb!]
\centering
\subfigure[Deceleration parameter for a closed universe.\label{qRk1}]
{\includegraphics[height=0.32\hsize,clip]{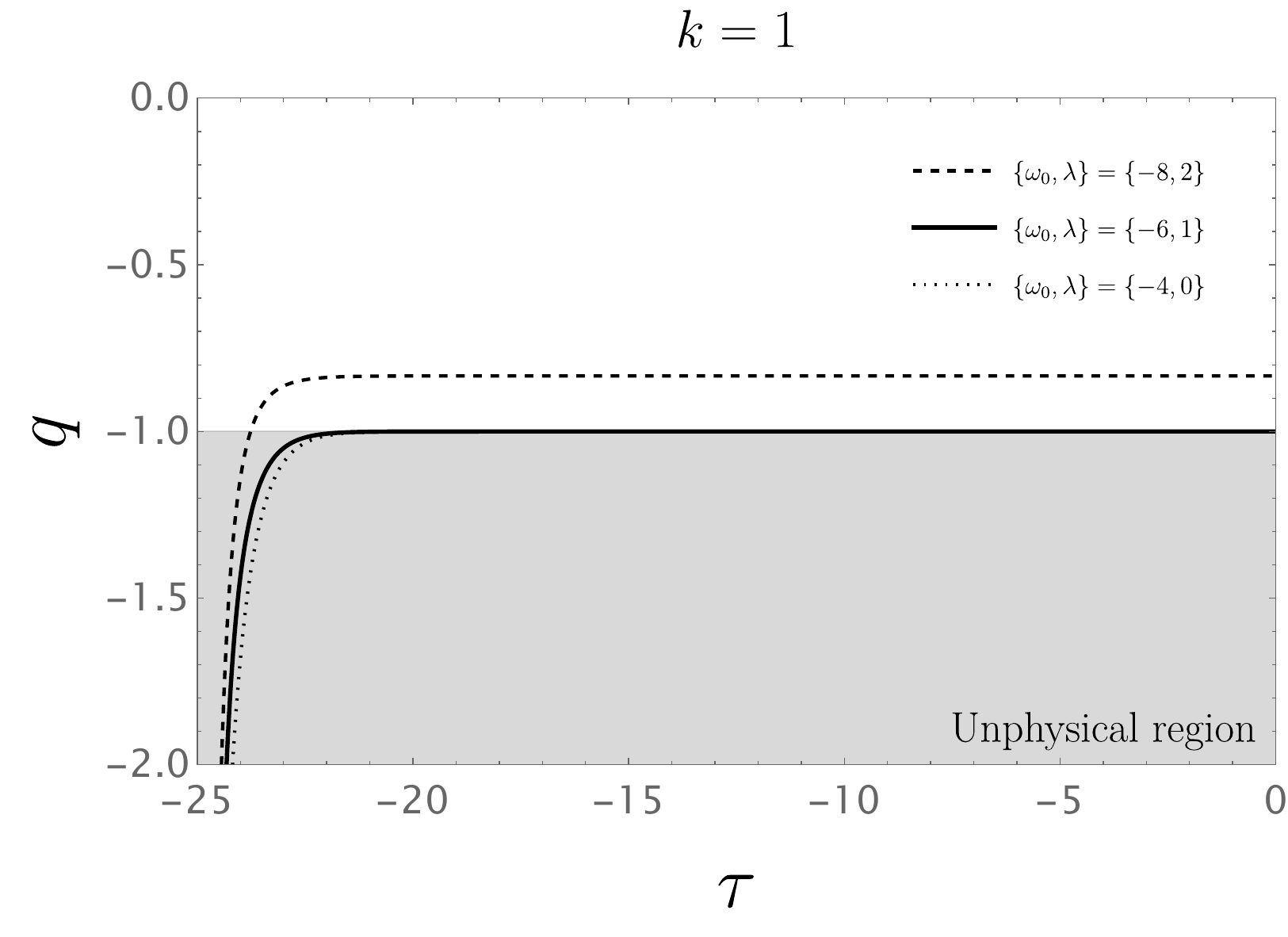}}
\subfigure[Deceleration parameter for an open universe.]
{\includegraphics[height=0.32\hsize,clip]{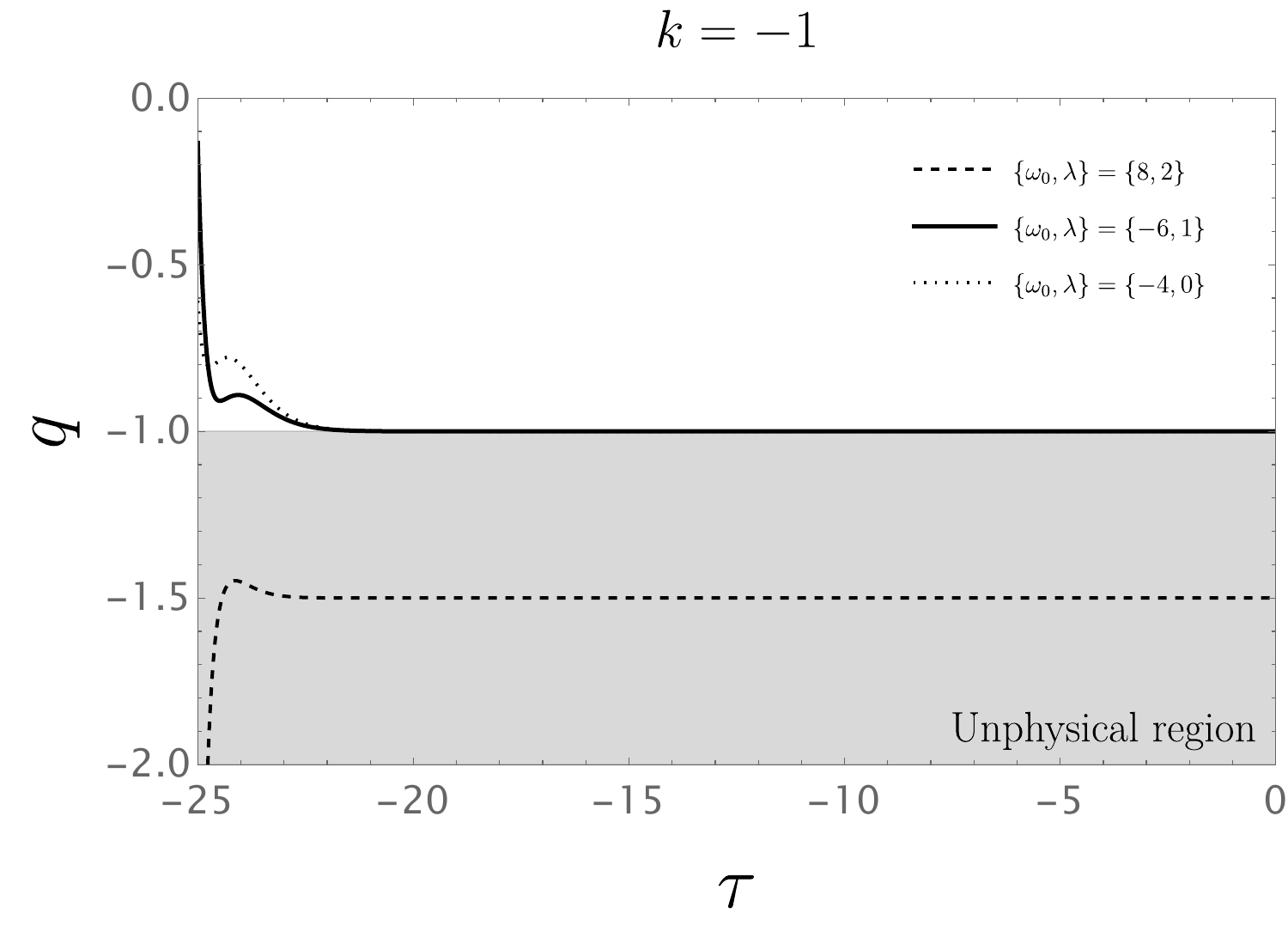}}
\caption{The deceleration parameters are computed under the assumption of a closed or an open universe in general relativity background. The initial conditions are deduced from the phase-space portrait diagrams. The initial conditions are selected as $\left(X_{\rm in}, \eta_{\rm in}\right) = \left(0, 0.2\right)$ for a closed universe, while they are provided by $\left(X_{\rm in}, \eta_{\rm in}\right) = \left(0.7, 0.7\right)$  for an open universe.}
\label{fig:qR}
\end{figure*}

At this stage, the introduction of compact variables is useful for constructing phase-space graphs.

We identified the stationary points within the finite regime. However, due to the possibility that the dynamical variable $x$ diverges, additional stationary points may arise at infinity. To investigate this, we introduce the compactified variable $X = \tanh{x}$, which maps the infinite domain of $x$ to a finite interval. The second variable, $\eta$, is already bounded by definition.

By selecting the free parameters appropriately, we construct the phase-space diagrams that illustrate the dynamical evolution of the system with the critical points explicitly identified in Fig.~\ref{fig:PSR}. Here, we observe that attractor points occur only when the spatial curvature is zero, i.e., $\eta=\pm 1$. This suggests that, regardless of whether the universe starts as closed or open, the dynamics of the scalar field drive it toward a flat FRW metric at late time. However, only the solution in which $\eta=1$ is physically acceptable if we want to interpret the scalar field as dark energy.

The phase-space diagrams also serve as a tool for determining the initial conditions required to solve the dynamical system and to analyze the properties of dilaton cosmology. Once the system is solved, we compute the deceleration parameter for both closed and open universes, with the results presented in Figs.~\ref{fig:qR}. The choice of free parameters for the phase-space diagrams follows from the stability analysis presented in Tab.~\ref{tab:EigenR}. In particular, the parameters are chosen to encompass as many critical points as possible in order to detect the whole dynamic behavior of the scalar field.

\begin{itemize}
    \item[-] {\bf Closed universe:}  In the case of a closed universe, the phase-space graphs are performed by selecting $\{\omega_{0},\lambda\}=\{-8,2\}$, $\{\omega_{0},\lambda\}=\{-6,1\}$, and $\{\omega_{0},\lambda\}=\{-4,0\}$ as free parameters, as shown in Fig.\ref{fig:PSR}. For these choices, three attractor points emerge according to the stability analysis presented in Tab. \ref{tab:EigenR}, namely $P_{R,0}^{k=1}$, $P_{R,5}^{k=1}$, and $P_{R,7}^{k=1}$. By appropriately choosing the initial conditions to solve the autonomous system, the dynamics of the scalar field evolve toward these critical points.

However, an analysis of the deceleration parameter reveals that only the attractor point $P_{R,0}^{k=1}$ corresponds to a physically viable scenario of late time accelerated expansion. This result is also supported by the fact that at the other critical points $\eta = -1$, which indicates a contracting universe at late times.

The behavior of the deceleration parameter for the dilaton field is given by the Fig.~\ref{fig:qR}, where the initial conditions are set as $\left(X_{\rm in}, \eta_{\rm in}\right)=\left(0, 0.2\right)$ and the dynamical system converges to the attractor point $P_{R,0}^{k=1}$. Here, we can observe that the deceleration parameter initially lies within the forbidden unphysical region, and for the parameter sets $\left\{\omega_0, \lambda\right\} = \left\{-6, 1\right\}$ and $\left\{\omega_0, \lambda\right\} = \left\{-4, 0\right\}$, the scalar field effectively mimics a cosmological constant at late times.

\item[-] {\bf Open universe:}
The phase-space structure of an open universe exhibits key differences compared to the closed case. Specifically, the critical points $P_{R,2}^{k=-1}$ and $P_{R,3}^{k=-1}$ exhibit different eigenvalues with respect to the closed case, while two additional critical points, $P_{R,8}^{k=-1}$ and $P_{R,9}^{k=-1}$, emerge in the dynamical system. These modifications lead to a different evolution of the system, impacting the late time attractor solutions.

To illustrate these differences, Fig.~\ref{fig:PSR} presents the phase-space graphs for selected values of the free parameters, namely $\{\omega_{0},\lambda\}=\{8,2\}$, $\{\omega_{0},\lambda\}=\{-6,1\}$, and $\{\omega_{0},\lambda\}=\{-4,0\}$. The analysis of these graphs reveals that the attractor points vary depending on the chosen parameter set.

For $\{\omega_{0},\lambda\}=\{8,2\}$, the system approaches the attractor points $P_{R,0}^{k=-1}$ and $P_{R,2}^{k=-1}$. In contrast, for $\{\omega_{0},\lambda\}=\{-6,1\}$ and $\{\omega_{0},\lambda\}=\{-4,0\}$, the attractor points are given by $P_{R,0}^{k=-1}$, $P_{R,5}^{k=-1}$, and $P_{R,7}^{k=-1}$.

Consequently, the behavior of the deceleration parameter also deviates from that of the open universe case. As shown in Fig.~\ref{fig:qR}, where we consider $\left(X_{\rm in},\eta_{\rm in}\right)=\left(0.7,0.7\right)$, the only valid attractor point, in which the universe undergoes a phase of accelerated expansion across all the considered free parameter values, is again $P_{R,0}^{k=-1}$. Indeed, the other points exhibit $\eta=-1$, indicating a contracting universe instead of an accelerated one.
Therefore, to interpret the scalar field as dark energy, only the attractor point $P_{R,0}^{k=-1}$ serves as a viable physical solution.
In Fig.~\ref{fig:qR}, we also observe that the case with $\left\{\omega_0, \lambda\right\} = \left\{-4, 0\right\}$ appears disfavored compared to the others, as the deceleration parameter never crosses the unphysical region. In contrast, for $\left\{\omega_0, \lambda\right\} = \left\{-6, 1\right\}$ and $\left\{\omega_0, \lambda\right\} = \left\{-8, 2\right\}$, the deceleration parameter remains within a physically acceptable range throughout the evolution.

Moreover, the open universe scenario appears to be favored over the closed one, since for the parameter sets $\left\{\omega_0, \lambda\right\} = \left\{-6, 1\right\}$ and $\left\{\omega_0, \lambda\right\}=\left\{-8, 2\right\}$, the deceleration parameter exhibits viable behavior across the entire cosmic history.
\end{itemize}

\subsection{Analysis with torsion}

For the teleparallel theory, the dimensionless variables used to analyze the stability of the scalar field are the same as those in general relativity background. Thus, we apply the same method within this alternative gravitational framework, and the corresponding eigenvalues are presented in Tab.~\ref{tab:EigenT} for both closed and open universes. As in the previous case, the stability of the dynamical system is determined by the choice of free parameters $\omega_{0}$ and $\lambda$. The results are depicted in Fig.~\ref{fig:stabcond2}, where the critical points are classified as stable, saddle, or unstable, depending on the region of the $\{\omega_{0},\lambda\}$ parameter space.

\begin{table*}[htb!]
\begin{center}
\setlength{\tabcolsep}{1.2em}
\renewcommand{\arraystretch}{1.8}
\begin{tabular}{|c|c|}
\hline
\hline
{\bf Critical point} & {\bf Eigenvalues} \\
\hline\hline
\cline{1-2}
\hline\hline
$P_{T,0}^{k=\pm1}$ & $\left\{-\frac{2 (\lambda  (\lambda +2)+\omega_0)}{\omega_0},-\frac{(\lambda +2)^2}{\omega_0}-3\right\}$ \\
$P_{T,1}^{k=\pm1}$ & $\left\{\frac{2 (\lambda  (\lambda +2)+\omega_0)}{\omega_0},\frac{(\lambda +2)^2}{\omega_0}+3\right\}$\\
$P_{T,2}^{k=1}$ & $\left\{-\frac{\omega_0 (-\lambda  (2 \lambda +\omega_0+8)+\omega_0+20)-A \sqrt{\omega_0}}{\omega_0 (-2 \lambda -\omega_0)^{3/2}},-\frac{A \sqrt{\omega_0}+\omega_0 (-\lambda  (2 \lambda +\omega_0+8)+\omega_0+20)}{\omega_0 (-2 \lambda -\omega_0)^{3/2}}\right\}$ \\
$P_{T,3}^{k=1}$ & $\left\{\frac{\omega_0 (-\lambda  (2 \lambda +\omega_0+8)+\omega_0+20)-A \sqrt{\omega_0}}{\omega_0 (-2 \lambda -\omega_0)^{3/2}},\frac{A \sqrt{\omega_0}+\omega_0 (-\lambda  (2 \lambda +\omega_0+8)+\omega_0+20)}{\omega_0 (-2 \lambda -\omega_0)^{3/2}}\right\}$ \\
$P_{T,2}^{k=-1}$ & $\left\{\frac{-\left((\lambda -1) \sqrt{\omega_0}\right)-i \sqrt{(4 \lambda +\omega_0) (3 \lambda  (\lambda +2)+4 \omega_0)-\omega_0}}{\sqrt{\omega_0} \sqrt{2 \lambda  (\lambda +1)+\omega_0}},\frac{-(\lambda -1) \sqrt{\omega_0}+i \sqrt{(4 \lambda +\omega_0) (3 \lambda  (\lambda +2)+4 \omega_0)-\omega_0}}{\sqrt{\omega_0} \sqrt{2 \lambda  (\lambda +1)+\omega_0}}\right\}$ \\
$P_{T,3}^{k=-1}$ & $\left\{\frac{(\lambda -1) \sqrt{\omega_0}-i \sqrt{(4 \lambda +\omega_0) (3 \lambda  (\lambda +2)+4 \omega_0)-\omega_0}}{\sqrt{\omega_0} \sqrt{2 \lambda  (\lambda +1)+\omega_0}},\frac{(\lambda -1) \sqrt{\omega_0}+i \sqrt{(4 \lambda +\omega_0) (3 \lambda  (\lambda +2)+4 \omega_0)-\omega_0}}{\sqrt{\omega_0} \sqrt{2 \lambda  (\lambda +1)+\omega_0}}\right\}$\\
$P_{T,4}^{k=\pm1}$ & $\left\{-4+\frac{4 i \sqrt{3}}{\sqrt{\omega_0}},-6+\frac{2 i \sqrt{3} (\lambda +2)}{\sqrt{\omega_0}}\right\}$\\
$P_{T,5}^{k=\pm1}$ & $\left\{4-\frac{4 i \sqrt{3}}{\sqrt{\omega_0}},6-\frac{2 i \sqrt{3} (\lambda +2)}{\sqrt{\omega_0}}\right\}$  \\
$P_{T,6}^{k=\pm1}$ & $\left\{-4-\frac{4 i \sqrt{3}}{\sqrt{\omega_0}},-6-\frac{2 i \sqrt{3} (\lambda +2)}{\sqrt{\omega_0}}\right\}$ \\
$P_{T,7}^{k=\pm1}$ & $\left\{4+\frac{4 i \sqrt{3}}{\sqrt{\omega_0}},6+\frac{2 i \sqrt{3} (\lambda +2)}{\sqrt{\omega_0}}\right\}$ \\
$P_{T,8}^{k=1}$ & $\left\{-\frac{2 i (\omega_0+3)}{\sqrt{3} \sqrt{\omega_0}},\frac{2 i (3 \lambda +\omega_0)}{\sqrt{3} \sqrt{\omega_0}}\right\}$ \\
$P_{T,9}^{k=1}$ & $\left\{\frac{2 i (\omega_0+3)}{\sqrt{3} \sqrt{\omega_0}},-\frac{2 i (3 \lambda +\omega_0)}{\sqrt{3} \sqrt{\omega_0}}\right\}$ \\
$P_{T,8}^{k=-1}$ & $\left\{\frac{2 (\omega_0+3)}{\sqrt{\omega_0 (2 \omega_0+3)}},-\frac{2 (3 \lambda +\omega_0)}{\sqrt{\omega_0 (2 \omega_0+3)}}\right\}$  \\

$P_{T,9}^{k=-1}$ & $\left\{-\frac{2 (\omega_0+3)}{\sqrt{\omega_0 (2 \omega_0+3)}},\frac{2 (3 \lambda +\omega_0)}{\sqrt{\omega_0 (2 \omega_0+3)}}\right\}$ \\
\hline\hline

\end{tabular}
\caption{The eigenvalues for the scalar field system in teleparallel gravity with exponential potential are given along with their stability conditions. Here, the parameter $A$ is given by $A=\sqrt{((\lambda -18) \lambda -55) \omega_0^3+4 (\lambda +1) ((\lambda -4) \lambda -54) \omega_0^2+4 (\lambda  (\lambda  (\lambda  (\lambda +4)-52)-192)-60) \omega_0-384 \lambda  (\lambda +2)-4 \omega_0^4}$, and the stability analysis for all the critical points is performed in Fig. \ref{fig:stabcond2}.}\label{tab:EigenT}
\end{center}
\end{table*}

\begin{figure*}[htb!]
\centering
\subfigure[Phase-space portrait graph for $\omega_{0}=-8$ and $\lambda=-2$.\label{1PSTk1}]
{\includegraphics[height=0.3\hsize,clip]{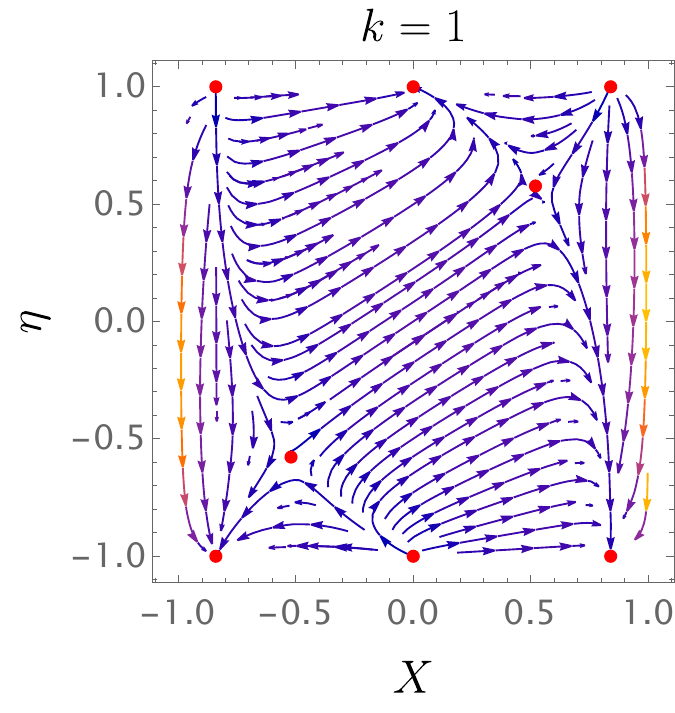}}
\subfigure[Phase-space portrait graph for $\omega_{0}=-3$ and $\lambda=-1$.\label{2PSTk1}]
{\includegraphics[height=0.3\hsize,clip]{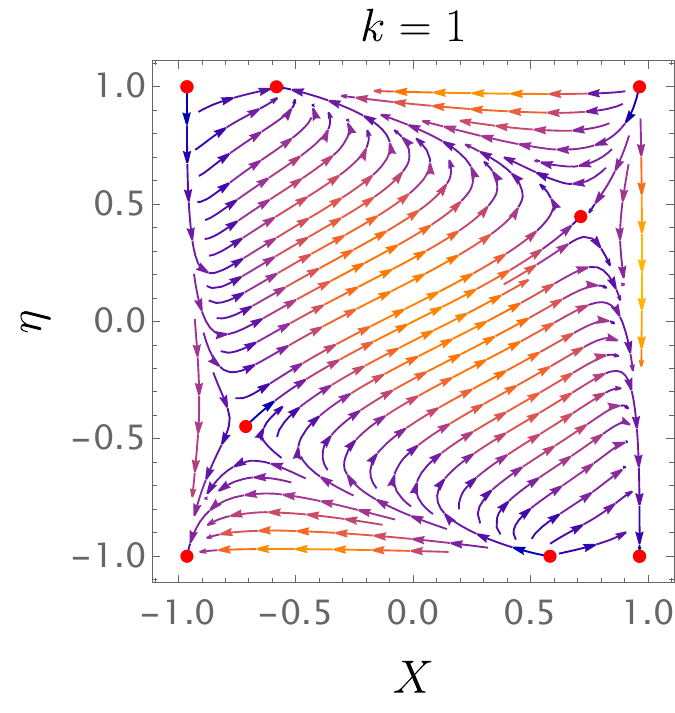}}
\subfigure[Phase-space portrait graph for $\omega_{0}=4$ and $\lambda=0$.\label{3PSTk1}]
{\includegraphics[height=0.3\hsize,clip]{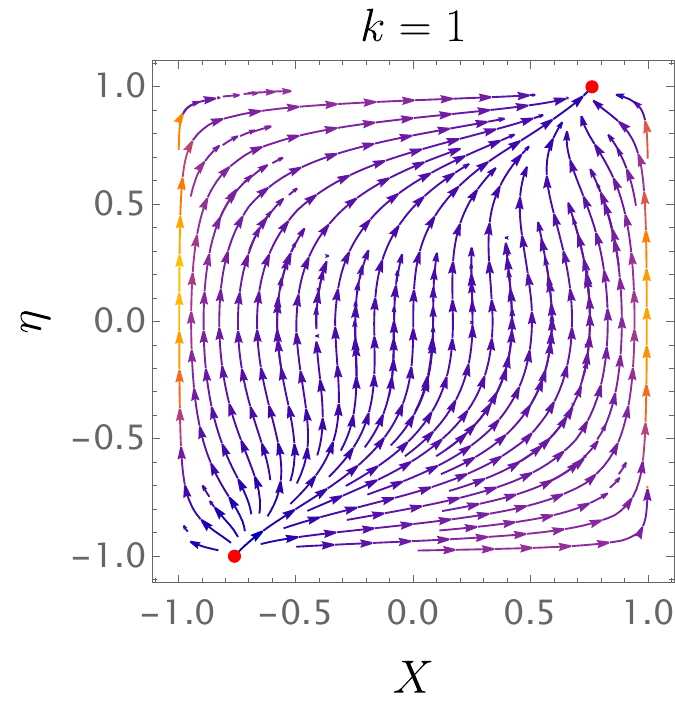}}
\centering
\subfigure[Phase-space portrait graph for $\omega_{0}=-8$ and $\lambda=-2$.\label{1PSTk-1}]
{\includegraphics[height=0.3\hsize,clip]{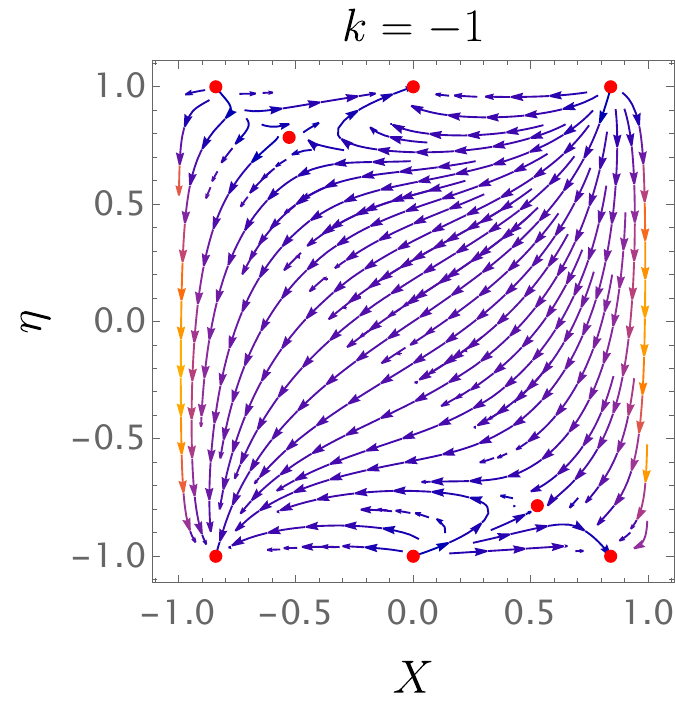}}
\subfigure[Phase-space portrait graph for $\omega_{0}=4$ and $\lambda=1$.\label{2PSTk-1}]
{\includegraphics[height=0.3\hsize,clip]{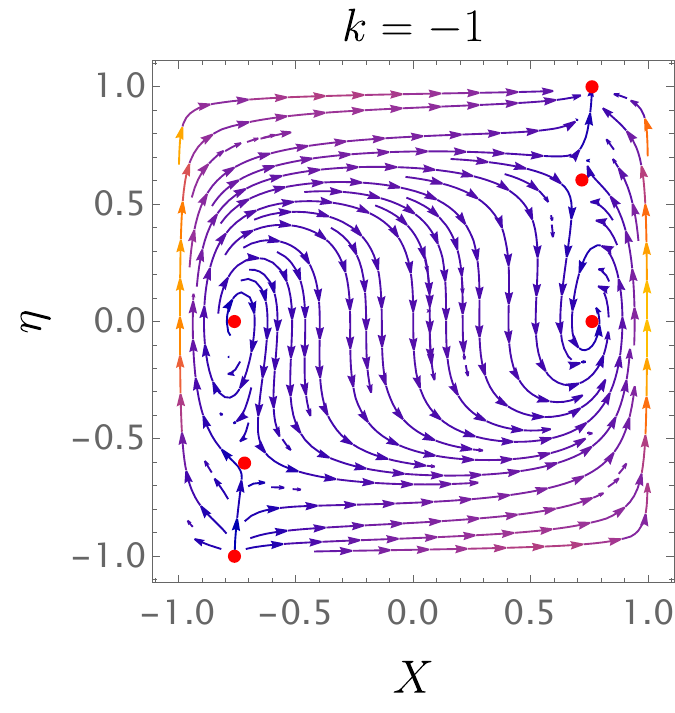}}
\subfigure[Phase-space portrait graph for $\omega_{0}=4$ and $\lambda=0$.\label{3PSTk-1}]
{\includegraphics[height=0.3\hsize,clip]{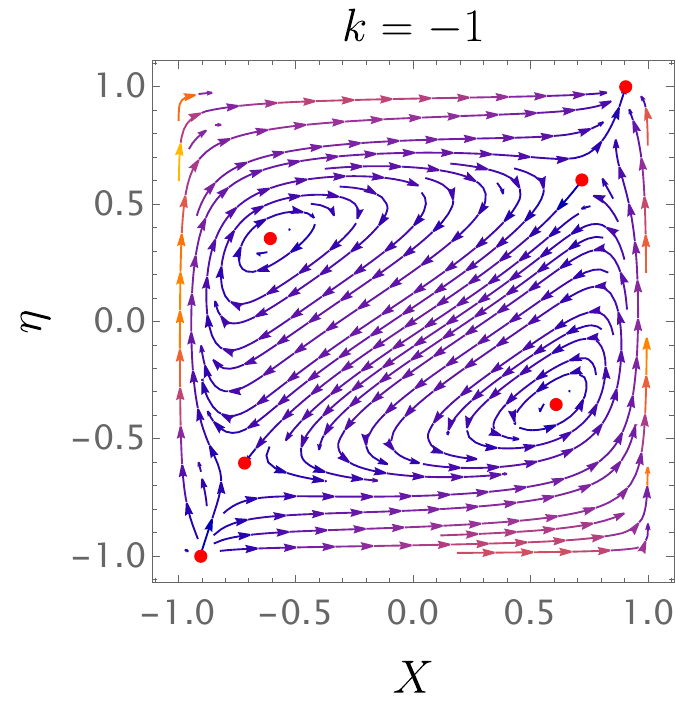}}
\caption{Phase-space portrait graphs for scalar field with the deceleration parameter in teleparallel gravity framework.}
\label{fig:PST}
\end{figure*}

\begin{figure*}[htb!]
\centering
\subfigure[Deceleration parameter for a closed universe.\label{qTk1}]
{\includegraphics[height=0.32\hsize,clip]{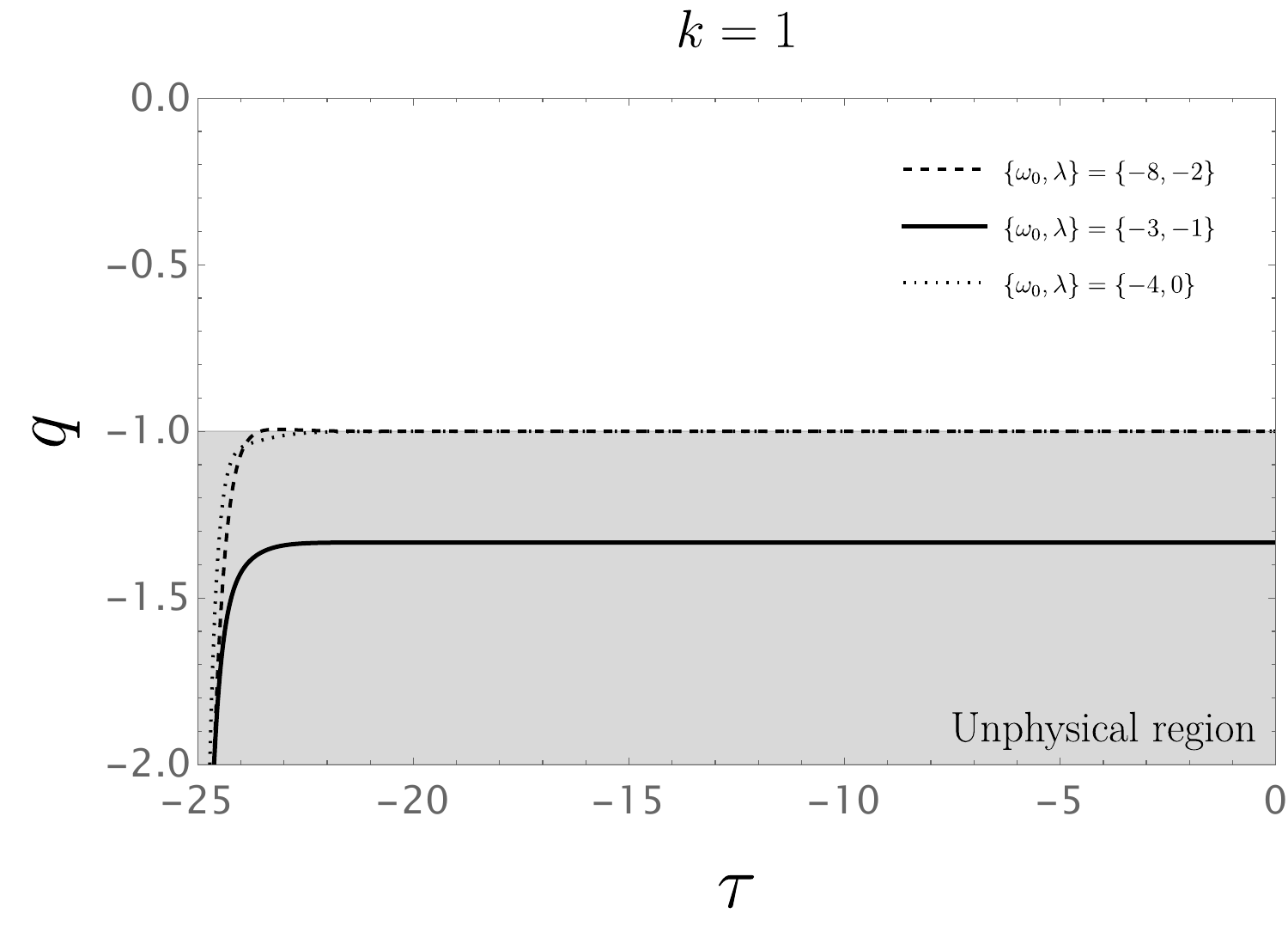}}
\subfigure[Deceleration parameter for an open universe.]
{\includegraphics[height=0.32\hsize,clip]{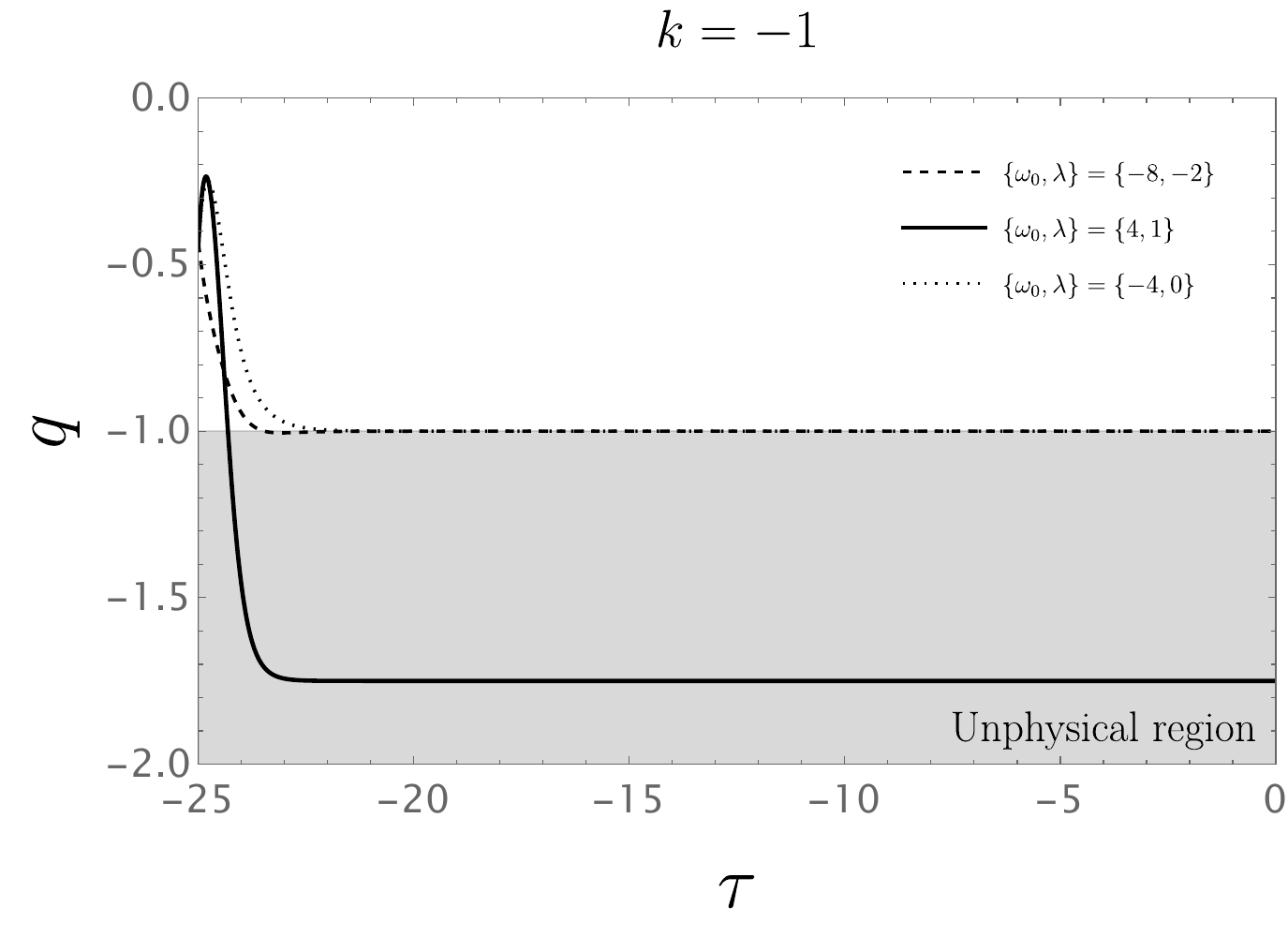}}
\caption{The deceleration parameters are computed under the assumption of a closed or an open universe in teleparallel background. The initial conditions are deduced from the phase-space portrait diagrams. The initial conditions are selected as $\left(X_{\rm in}, \eta_{\rm in}\right) = \left(-0.5, 0.5\right)$ for a closed universe, while they are provided by $\left(X_{\rm in}, \eta_{\rm in}\right) = \left(0, 0.8\right)$  for an open universe.}
\label{fig:qT}
\end{figure*}

Using these stability conditions, we construct the phase-space diagrams for both closed and open universes, as shown in Fig.~\ref{fig:PST}. These diagrams are also used to determine the initial conditions required to solve the dynamical system such that its evolution converges to a stable point, i.e., an attractor for the scalar field. Again, we select the free parameters to visualize the whole dynamics and observe the behavior of all critical points, according to Fig.~\ref{fig:stabcond2}.

\begin{itemize}
    \item[-]{\bf Closed universe:} For a closed universe, we consider the parameter sets $\left\{\omega_0, \lambda\right\} = \left\{-8, -2\right\}$, $\left\{\omega_0, \lambda\right\} =\left\{-3, -1\right\}$, and $\left\{\omega_0, \lambda\right\} =\left\{4, 0\right\}$. The attractor points of the system are $P_{T,0}^{k=1}$, $P_{T,4}^{k=1}$, and $P_{T,6}^{k=1}$ for $\left\{\omega_0, \lambda\right\} = \left\{-8, -2\right\}$ and $\left\{\omega_0, \lambda\right\} =\left\{-3, -1\right\}$, while for $\left\{\omega_0, \lambda\right\} =\left\{4, 0\right\}$, the only attractor is $P_{T,0}^{k=1}$, consistent with the condition $\omega_0 > 0$. As in the curvature background, the critical points $P_{T,4}^{k=1}$ and $P_{T,6}^{k=1}$ do not correspond to physically viable attractors, since $\eta = -1$ in these cases, indicating a collapsing universe. Consequently, the only attractor point where the scalar field behaves as an effective dark energy component at late times is $P_{T,0}^{k=1}$.

This conclusion is supported by the analysis of the deceleration parameter in Fig.~\ref{fig:qT}, where we set the initial conditions as $\left(X_{\rm in}, \eta_{\rm in}\right) = (-0.5, 0.5)$ to ensure that the dynamical system converges to $P_{T,0}^{k=1}$. From this figure, it is evident that the case $\left\{\omega_0, \lambda\right\} = \left\{-3, -1\right\}$ is ruled out, as it results in a persistent unphysical behavior throughout the cosmic evolution. In contrast, the scenarios with $\left\{\omega_0, \lambda\right\} = \left\{-8, -2\right\}$ and $\left\{\omega_0, \lambda\right\} = \left\{4, 0\right\}$ exit the unphysical region at late times, effectively mimicking the characteristics of a cosmological constant.

    \item[-]{\bf Open universe:}
For an open universe, we select the free parameters as $\{\omega_{0},\lambda\}=\{-8,-2\}$, $\{\omega_{0},\lambda\}=\{4,1\}$, and $\{\omega_{0},\lambda\}=\{4,0\}$. Under these conditions, the attractor points for $\{\omega_{0},\lambda\}=\{-8,-2\}$ are the same as in the closed case, i.e.,  $P_{T,1}^{k=-1}$, $P_{T,4}^{k=-1}$, and $P_{T,6}^{k=-1}$. Meanwhile, for $\{\omega_{0},\lambda\}=\{4,1\}$ and $\{\omega_{0},\lambda\}=\{4,0\}$, the attractor points are $P_{T,1}^{k=-1}$ and $P_{T,3}^{k=-1}$.

However, the only viable attractor point for the system is $P_{T,0}^{k=-1}$, while $P_{T,3}^{k=-1}$, $P_{T,4}^{k=-1}$, and $P_{T,6}^{k=-1}$ are ruled out. Specifically, $P_{T,4}^{k=-1}$ and $P_{T,6}^{k=-1}$ correspond to a collapsing universe since $\eta=-1$, whereas $P_{T,3}^{k=-1}$ does not exhibit a physically acceptable dynamical behavior.

The deceleration parameter shown in Fig. \ref{fig:qT} is obtained by setting $\left(X_{\rm in},\eta_{\rm in}\right)=\left(0,0.8\right)$. Here, we can deduce that by choosing $\{\omega_{0},\lambda\}=\{4,1\}$ as free parameters for our model, the system falls into the unphysical region at late times.  In contrast, for $\left\{\omega_0, \lambda\right\} = \left\{-8, -2\right\}$ and $\left\{\omega_0, \lambda\right\}=\left\{4, 0\right\}$, the system avoids the unphysical region throughout the entire cosmic evolution. Moreover, at late times, the scalar field in these scenarios effectively mimics the behavior of a cosmological constant.

These findings confirm that the scalar field plays a fundamental role in driving cosmic acceleration when the initial conditions are chosen near the appropriate attractor point.

\end{itemize}

\subsection{Analysis with non-metricity}

In the symmetric-teleparallel  theory, the stability analysis performed to obtain phase-space diagrams differs from that in the other two frameworks. This is because we have an additional scalar field—specifically, the scalar field in the non-metricity scalar given by Eq.~\eqref{eq:Qscalar}. Therefore, we consider an additional dimensionless variable in the stability analysis, increasing the non-linearity of the autonomous system.

Moreover, due to the non-linearity of the system, we have replaced the dynamical variable $\eta$ with $\Omega_k$ to study the effect of spatial curvature in the dilaton cosmology.

For these reasons, the small perturbations around a critical point are provided by
\begin{eqnarray}
\left(
\begin{array}{c}
\delta x' \\
\delta \Omega_{k}' \\
\delta z'\\
\end{array}
\right) = {\mathcal J} \left(
\begin{array}{c}
\delta x \\
\delta \Omega_{k} \\
\delta z\\
\end{array}
\right) \,,
\label{eq:pert1}
\end{eqnarray}
where the Jacobian matrix $\mathcal{J}$, evaluated at the critical point $P_{c}$, takes the following form
\begin{eqnarray}
 \label{matJ1}
\mathcal{J}=\left( \begin{array}{ccc}
\frac{\partial x'}{\partial x}& \frac{\partial x'}{\partial \Omega_{k}}& \frac{\partial x'}{\partial z}\\
\frac{\partial \Omega_{k}'}{\partial x}& \frac{\partial \Omega_{k}'}{\partial \Omega_{k}}& \frac{\partial \Omega_{k}'}{\partial z}\\
\frac{\partial z'}{\partial x}& \frac{\partial z'}{\partial \Omega_{k}}& \frac{\partial z'}{\partial z}\\
\end{array} \right)_{P_{c}}\,.
\label{eq:J1}
\end{eqnarray}

However, when applying these variables to analyze the dynamical behavior of the scalar field within the non-metricity framework, it is not possible to establish the stability of the system in the same manner as in the previous scenarios. This limitation arises, as just discussed, from the strong nonlinearity inherent to the system.

\begin{table*}
\begin{center}
\scriptsize
\setlength{\tabcolsep}{0.2em}
\renewcommand{\arraystretch}{1.8}
\begin{tabular}{|c|c|c|c|c|c|c|}
\hline\hline
{\bf Critical point} & \multicolumn{3}{c|}{{\bf Eigenvalues}} & \multicolumn{3}{c|}{{\bf Stability}} \\
\cline{2-7}
 & $w_0=-1$ & $w_0=0$ & $w_0=1$ & $w_0=-1$ & $w_0=0$ & $w_0=1$ \\
\hline\hline
$P_{Q,0}$ & $\left\{\frac{4 \sqrt{2}}{3}+2,\frac{4 \sqrt{2}}{3},-\frac{4 \sqrt{2}}{3}\right\}$ & $\{4, -2, 2\}$ & $-$ & Saddle & Saddle & - \\
$P_{Q,1}$ & $\left\{\frac{4 \sqrt{2}}{3},-\frac{4 \sqrt{2}}{3},2-\frac{4 \sqrt{2}}{3}\right\}$ & $-$ & $-$ & Saddle & - & - \\
$P_{Q,2}/P_{Q,3}$ & $\left\{\sqrt{\frac{22}{17}}+1,-2,1-\sqrt{\frac{22}{17}}\right\}$ & $\left\{\sqrt{41}+1,1-\sqrt{41},-2\right\}$ & $\left\{2 \sqrt{79}+1,1-2 \sqrt{79},-2\right\}$ & Saddle & Saddle & Saddle \\
$P_{Q,4}$ & $-$ & $-$ & $-$ & $-$ & $-$ & $-$ \\
$P_{Q,5}$ & $-$ & $-$ & $-$ & $-$ & $-$ & $-$ \\
$P_{Q,6}$ & $\{2,1.4,1.4\}$ & $-$ & $\{2,-1.5,-1.5\}$ & Unstable & $-$ & Saddle \\
$P_{Q,7}$ & $\{2,-4.5,-4.5\}$ & $-$ & $\{2,-1.5,-1.5\}$ & Saddle & $-$ & Saddle \\
\hline\hline
\end{tabular}
\caption{The eigenvalues for the scalar field system imposing $\lambda=0$ in symmetric-teleparallel gravity with exponential potential are given along with their stability conditions for different values of $w_0$.}
\label{tab:EigenQ}
\end{center}
\end{table*}
Then, we restrict our analysis to specific and physically meaningful values of $\{\omega_0,\lambda\}$. In particular, we consider two limiting cases:

\begin{itemize}
    \item[-] When $\omega_0 \to 0$, the system reduces to a purely $f(Q)$ framework,
     \item[-] When $\lambda \to 0$, the potential remains constant, effectively mimicking the behavior of a cosmological constant.
\end{itemize}

Based on these considerations, we focus our study on the parameter sets with  $\omega_0 \in \left\{-1, 0, 1\right\}$ and $\lambda = 0$. This choice allows us to investigate the stability of the scalar field in the non-metricity framework, clarifying how variations in $\omega_0$ affect the cosmological constant-like scenario. The eigenvalues and the stability properties of the critical points are summarized in Tab.~\ref{tab:EigenQ}. We observe that, for the selected parameters, not all points qualify as critical points of the dynamical system. Specifically, the points $P_{Q,4}$ and $P_{Q,5}$ do not exist for $\lambda = 0$ and are therefore excluded from the stability analysis. Furthermore, setting $\omega_0 = 0$ eliminates the points $P_{Q,1}$, $P_{Q,5}$, $P_{Q,6}$, and $P_{Q,7}$, while choosing $\omega_0 = 1$ renders the points $P_{Q,0}$ and $P_{Q,1}$ unavailable. The results are not equivalent to those given in the torsion background, since we are working in the non-coincident gauge. For this choice, Eq. \eqref{eq:Qscalar} does not take the same value as Eq. \eqref{eq:Tscalar} due to the presence of a new additional field in the non-metricity scalar.

\section{Final remarks and perspectives}\label{sec5}

In this work, we investigated the role of a non-minimally coupled scalar field where the gravity sector is identified as either the Ricci, torsion or non-metric scalars, considered individually. The non-minimally coupled part, between the scalar field and the above cited scalars, is rewritten to resemble the form of a Brans-Dicke scalar field Lagrangian, in order to compactly reduce the complexity of our analysis, albeit leaving unaltered the physical properties of our dilaton-inspired paradigm.

Accordingly, in this dilaton-inspired scenario, the three distinct formulations of gravity appear, at first glance, dynamically equivalent, forming the so-called \emph{trinity of gravity}, recently widely considered in the literature, to describe gravitational interactions, without passing forcedly through the spacetime curvature only.

From the perspective of background, the three models may appear equivalent, as the equations of motion are constructed similarly. However, we explicitly showed that their behaviors appear quite different and significantly altered,  when the dilaton-inspired field is coupled to gravity via a non-minimal interaction.

To assess these differences, we performed a background-level stability analysis in each case, assuming an exponential potential for the scalar field. In order to catch relevant cosmological consequences, since the dilaton perspective is inspired by string theory, we specialized our scalar field framework, from the early universe onward and incorporated spatial curvature through a non-flat FRW metric, fulfilling the cosmological principle and guaranteeing the important feature of dilatons, i.e., providing a non-zero spatial curvature.

Our analysis definitively demonstrated that in both general relativity and torsion a complete linear stability study was feasible. In these settings, we identified critical points that act as attractors in the corresponding dynamical systems, with their existence and nature governed by the free parameters $\omega_0$ and $\lambda$.

In general relativity, we found that critical points existed for both positive and negative values of $\omega_0$, ensuring broad parametric viability. This flexibility, however, was not preserved in the teleparallel scenario, where most of the critical points allow only negative values of $\omega_0$, thereby favoring a phantom-like regime for the dilaton-inspired field.

Further, the stability analysis in these two frameworks revealed that, with suitable choices of the free parameters, the deceleration parameter approached values compatible with a cosmological constant at late times, for both closed and open spatial geometries.

Remarkably, involving as hypothesis, an open universe, appeared more robust, as it allows for smooth and well-behaved dynamical evolution across the entire cosmic expansion history.

Conversely, in the non-metricity framework, a complete stability analysis is hindered by the non-linearities introduced through the non-coincident gauge condition. Indeed, this gauge choice introduces an additional scalar field into the system, which requires a distinct stability analysis compared to those performed in the curvature and torsion-based backgrounds.

Within this perspective, we examined various regimes of the kinetic term, spanning from phantom-like to quintessence-like behaviors, although we found no attractor solutions. This suggested that the dilaton field in the non-metricity background was disfavored from dynamical stability and, meanwhile, the backgrounds cannot turn into an equivalent description, as commonly considered.

Overall, our results highlighted a remarkable property of the scalar field within both general relativity and teleparallel gravity: in each framework, the field demonstrates the capacity to account for the universe's late time acceleration in a dynamically stable manner. In particular, we observed that, for suitable choices of the free parameters, our scalar field can mimic a cosmological constant at late times, once the corresponding critical point is reached.

We finally interpreted our findings in view of cosmographic outcomes, singling out those regions that appear unphysical and remarking those, instead, showing viable characteristics to describe the late time dynamics, moving onward from early times.

Future works will shed light on computing higher-order cosmographic parameters, invoking a more accurate stability analysis. Moreover, the study of additional non-minimal couplings, to add inside the underlying Lagrangian, can be used to clarify how the stability description is dependent on the type of couplings. A more robust and physically motivated dilaton field can also be considered, by coupling the here adopted Lagrangian with a vector field, providing the presence of either photons or Proca-like terms. We can therefore expect to find a more accurate descriptions of the physical reasons that do not permit the third gravity-equivalent description, the non-metricity, to equivalently describe gravity, under stability analyses. This will be compared with previous literature in which similar results have also been found for quintessence-like and quasiquintessence fields \cite{Carloni:2024ybx}.


\begin{acknowledgments}
YC is grateful to Anna Chiara Alfano, Tommaso Mengoni and Alessandro Saltarelli for insightful discussions on affine subjects related to this paper. OL acknowledges support by the  Fondazione  ICSC, Spoke 3 Astrophysics and Cosmos Observations. National Recovery and Resilience Plan (Piano Nazionale di Ripresa e Resilienza, PNRR) Project ID $CN00000013$ ``Italian Research Center on  High-Performance Computing, Big Data and Quantum Computing" funded by MUR Missione 4 Componente 2 Investimento 1.4: Potenziamento strutture di ricerca e creazione di ``campioni nazionali di R\&S (M4C2-19)" - Next Generation EU (NGEU). AP thanks the support of VRIDT through Resoluci´on VRIDT No. 096/2022 and Resoluci´on VRIDT
No. 098/2022. Part of this study was supported by
FONDECYT 1240514 ETAPA 2025.
\end{acknowledgments}


\bibliographystyle{ieeetr}

\appendix

\section{Region plots for the stability analysis}\label{app}

In this appendix, we delineate the regions in the $\{\omega_{0}, \lambda\}$ parameter space where the critical points exhibit stable, saddle, or unstable behavior. A full stability analysis of the dilaton field is provided for both general relativity and teleparallel theories of gravity.

\begin{figure*}[htb!]
\centering
\subfigure[$\{\omega_0,\lambda\}$ region to identify the stability of $P_{R,0}^{k=\pm 1}$ and $P_{R,1}^{k=\pm 1}$\label{PR01}]
{\includegraphics[height=0.4\hsize,clip]{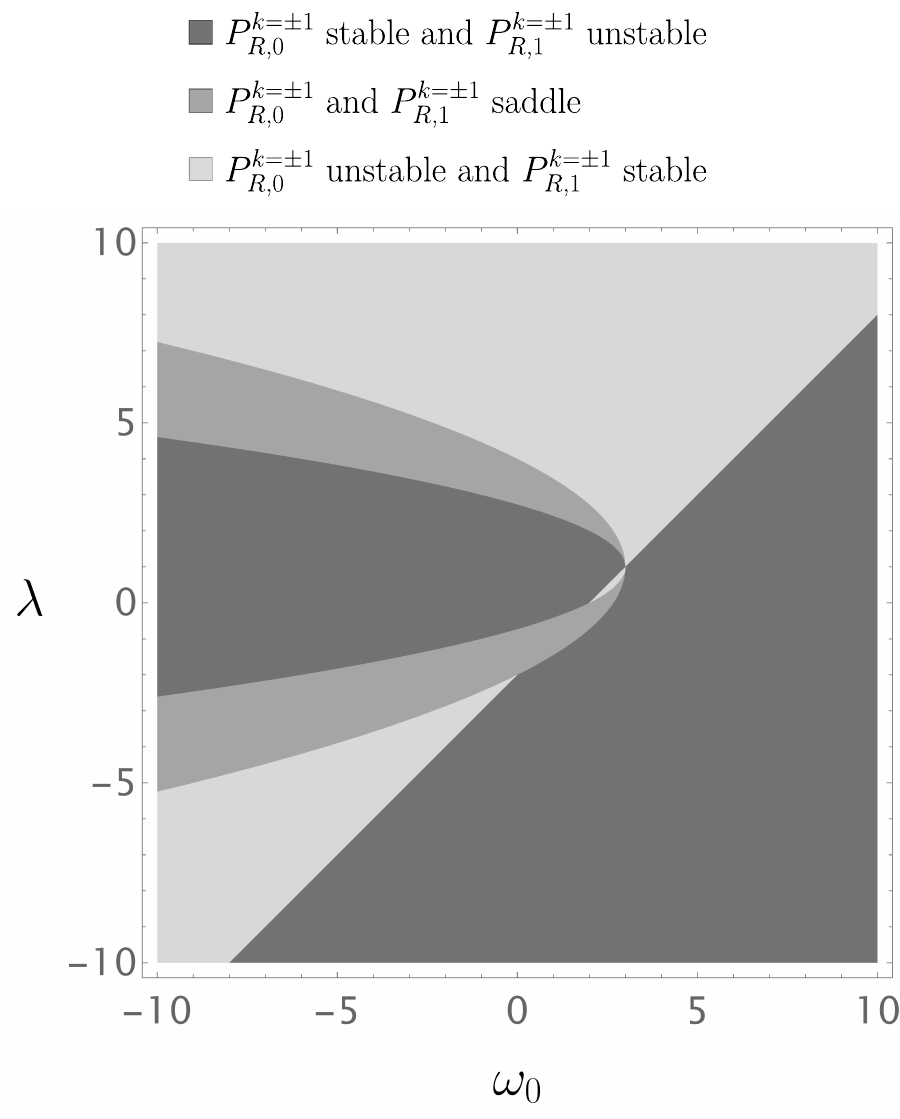}}
\subfigure[$\{\omega_0,\lambda\}$ region to identify the stability of $P_{R,4}^{k=\pm 1}$ and $P_{R,5}^{k=\pm 1}$.\label{PR45}]
{\includegraphics[height=0.4\hsize,clip]{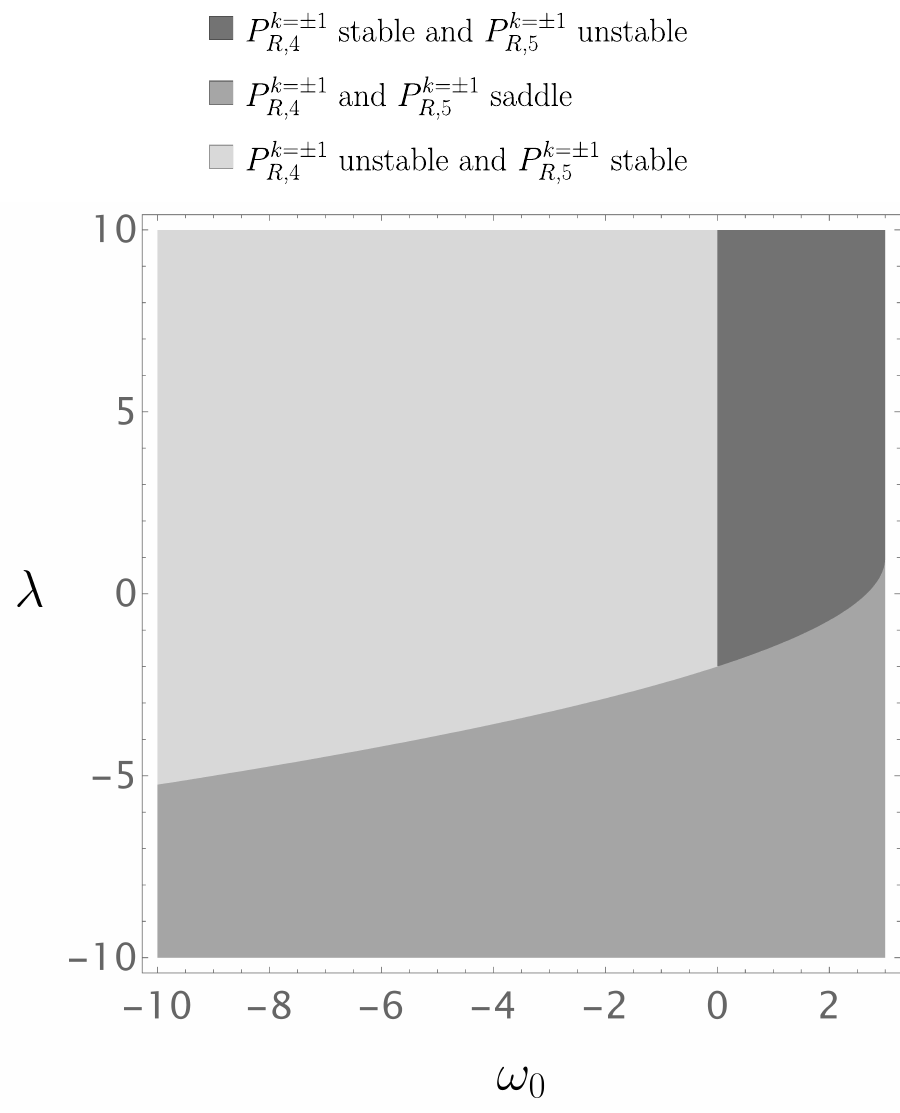}}
\subfigure[$\{\omega_0,\lambda\}$ region to identify the stability of $P_{R,6}^{k=\pm 1}$ and $P_{R,7}^{k=\pm 1}$.\label{PR67}]
{\includegraphics[height=0.4\hsize,clip]{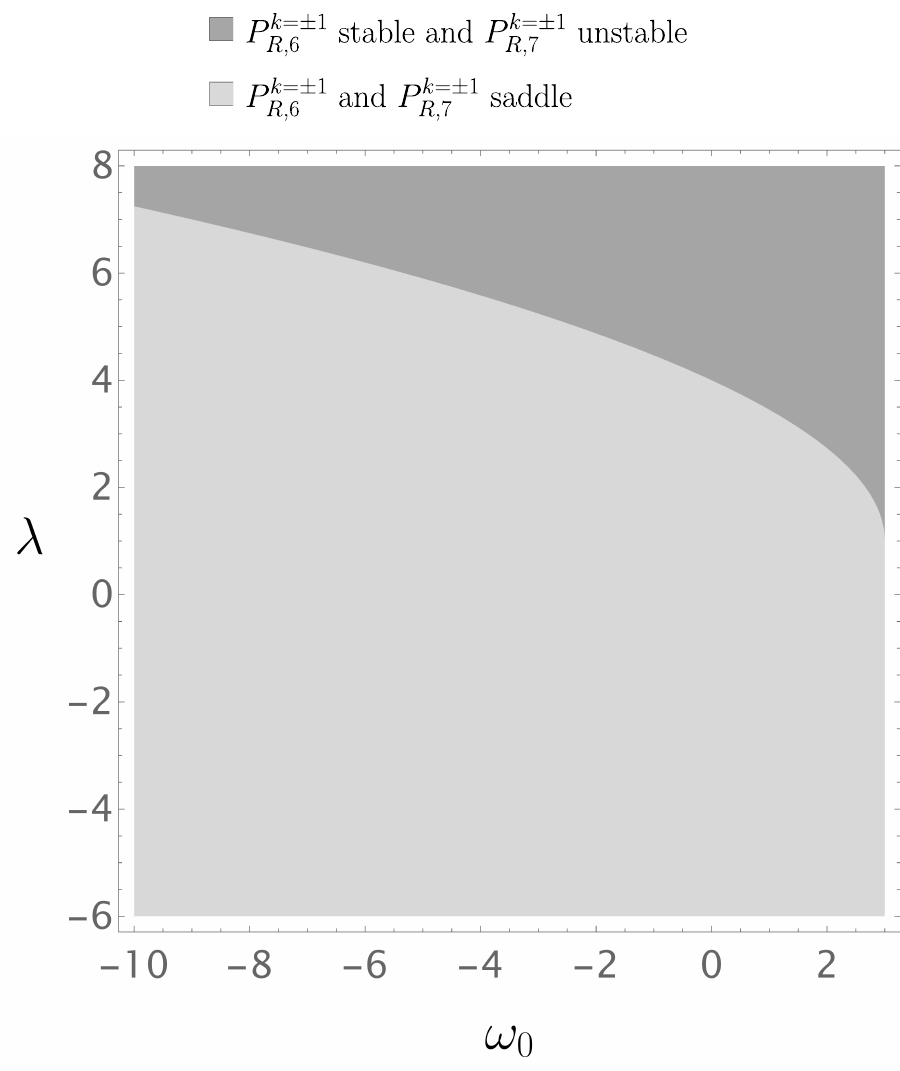}}
\subfigure[$\{\omega_0,\lambda\}$ region to identify the stability of $P_{R,2}^{k=-1}$ and $P_{R,3}^{k=-1}$.\label{PR23}]
{\includegraphics[height=0.4\hsize,clip]{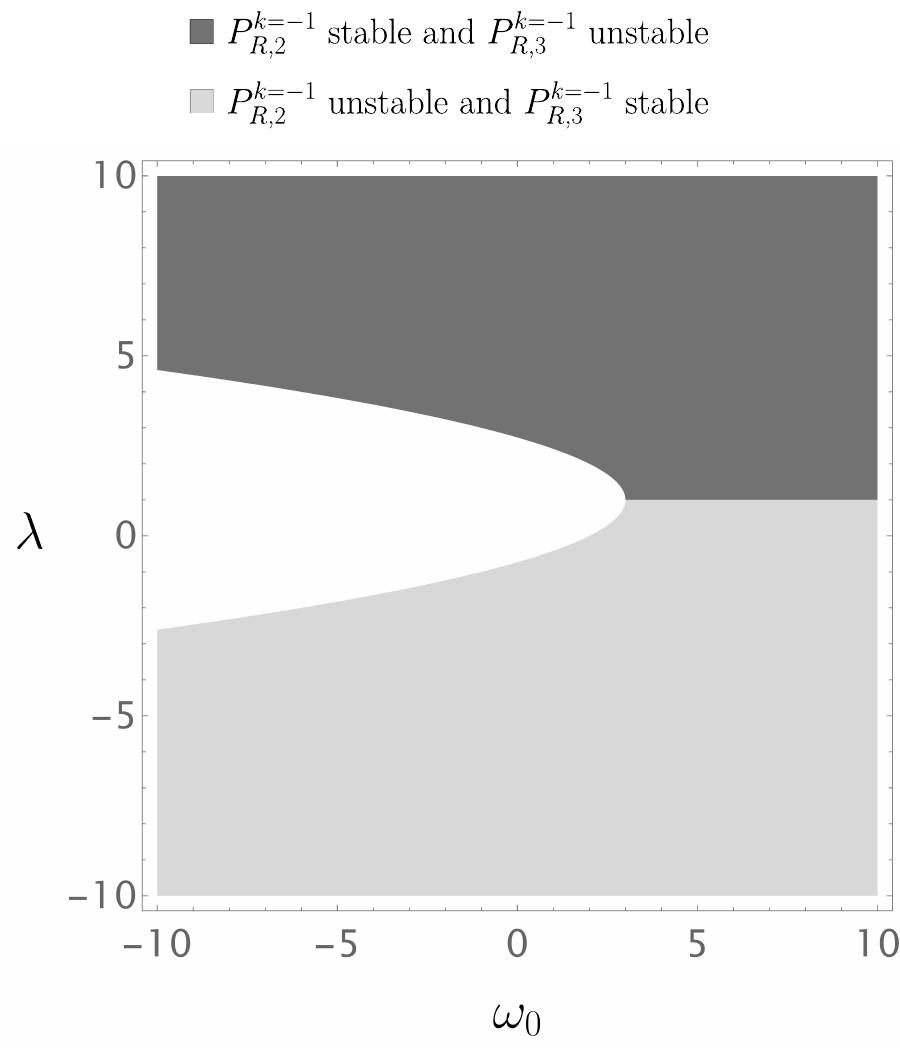}}
\caption{Free parameter regions to establish scalar field stability in general relativity for both closed and open universes.}
\label{fig:stabcond1}
\end{figure*}

\begin{figure*}[htb!]
\centering
\subfigure[$\{\omega_0,\lambda\}$ region to identify the stability of $P_{T,0}^{k=\pm 1}$ and $P_{T,1}^{k=\pm 1}$\label{PT01}]
{\includegraphics[height=0.4\hsize,clip]{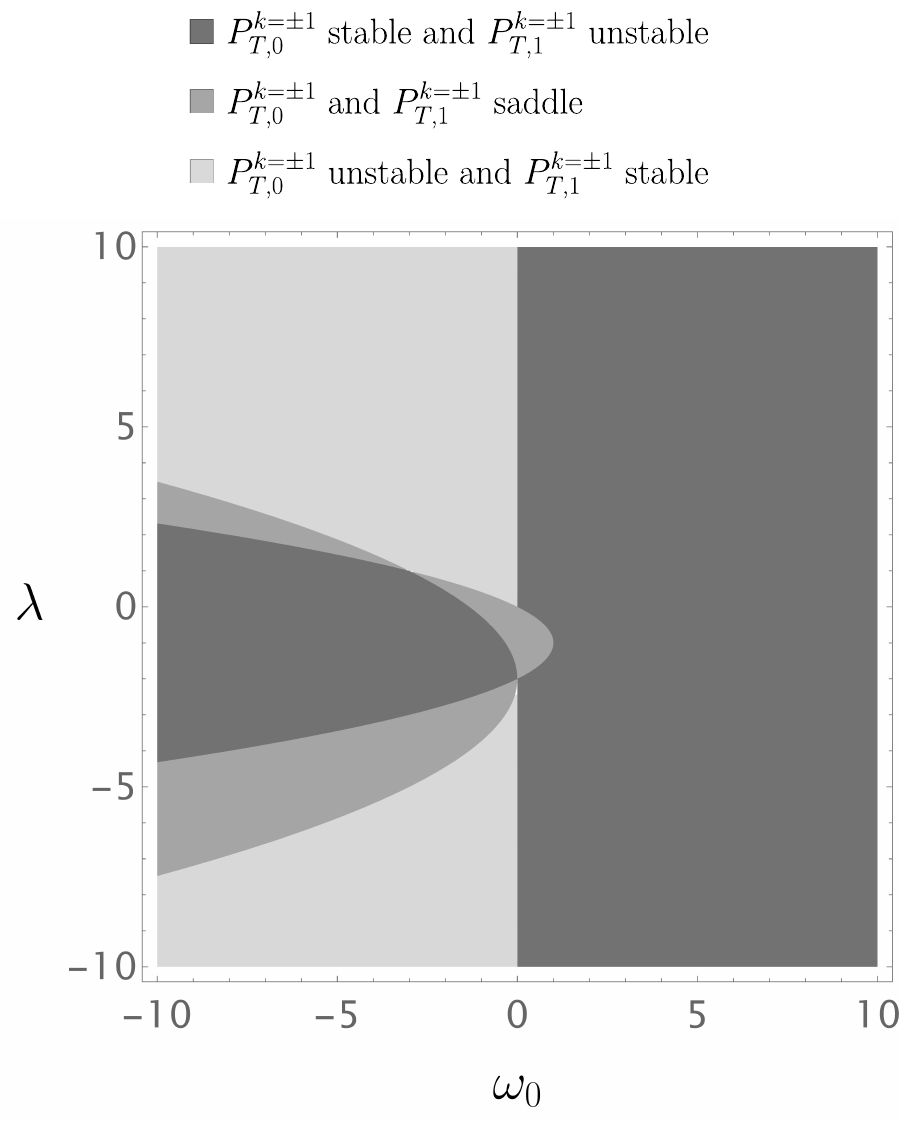}}
\subfigure[$\{\omega_0,\lambda\}$ region to identify the stability of $P_{T,2}^{k=1}$ and $P_{T,3}^{k= 1}$.\label{PT23}]
{\includegraphics[height=0.4\hsize,clip]{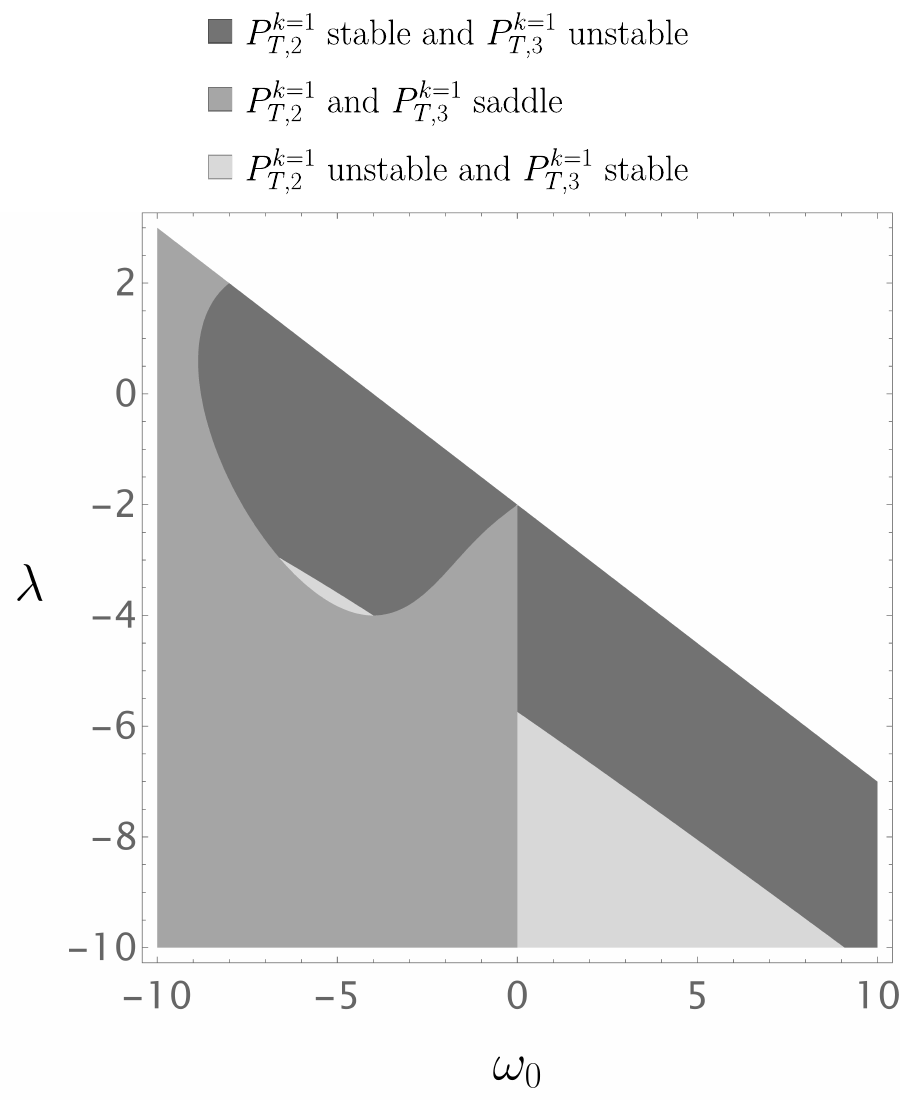}}
\subfigure[$\{\omega_0,\lambda\}$ region to identify the stability of $P_{T,2}^{k=-1}$ and $P_{T,3}^{k=-1}$.\label{PT23-1}]
{\includegraphics[height=0.4\hsize,clip]{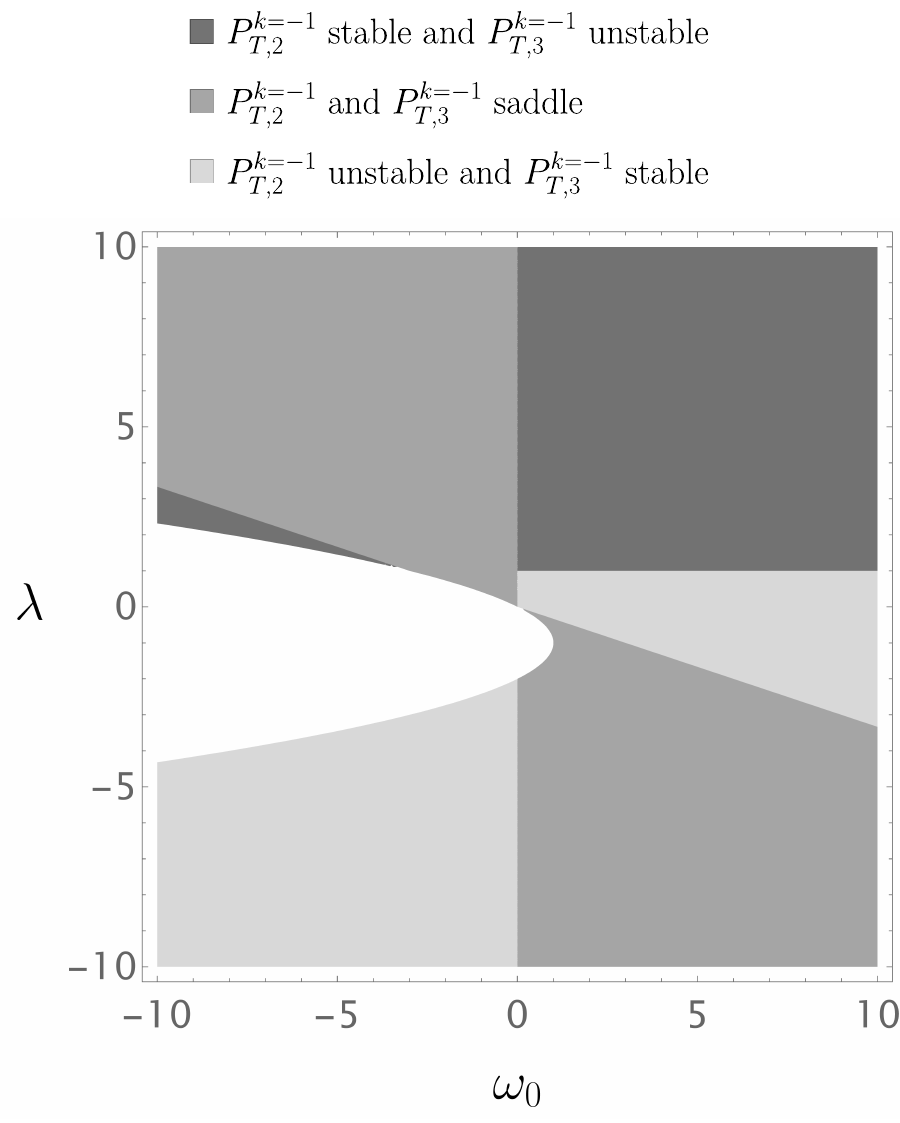}}
\subfigure[$\{\omega_0,\lambda\}$ region to identify the stability of $P_{T,4}^{k=\pm1}$ and $P_{T,5}^{k=\pm1}$.\label{PT45}]
{\includegraphics[height=0.4\hsize,clip]{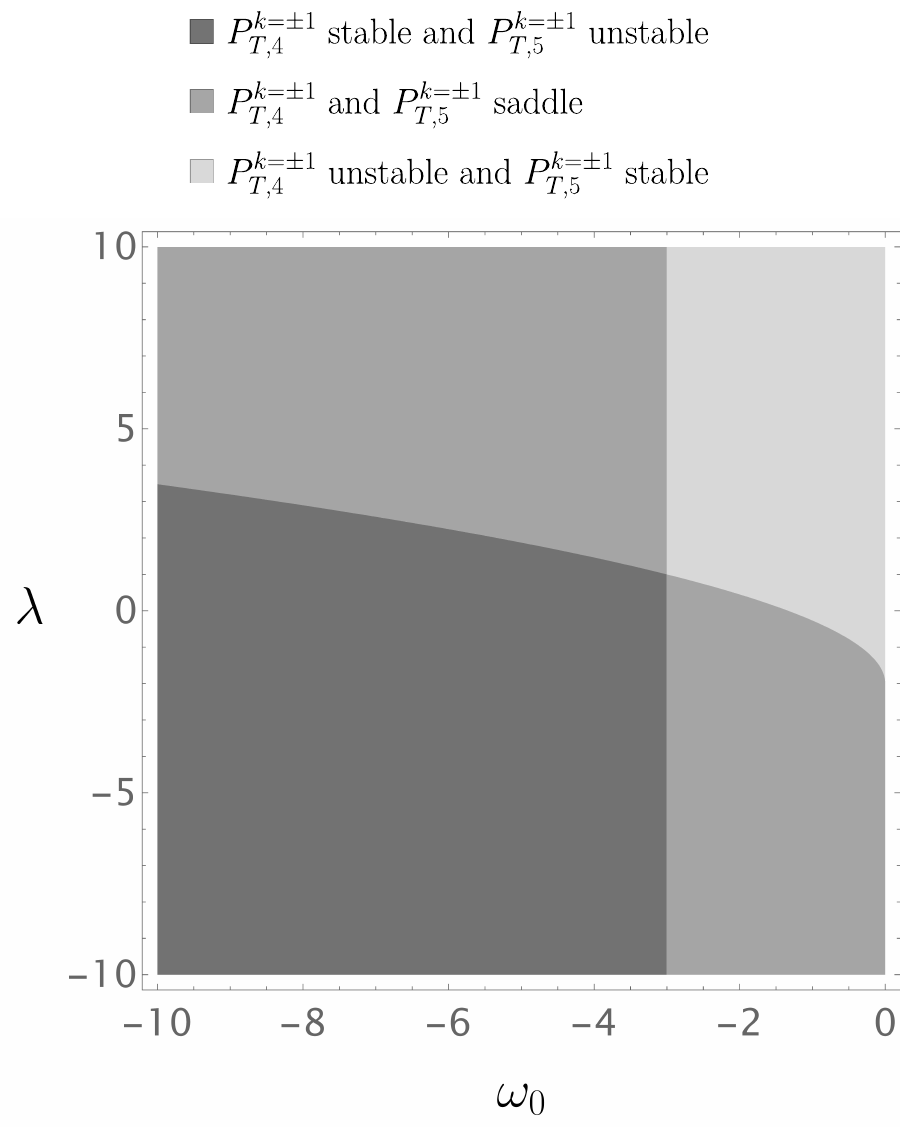}}
\subfigure[$\{\omega_0,\lambda\}$ region to identify the stability of $P_{T,6}^{k=\pm1}$ and $P_{T,7}^{k=\pm1}$.\label{PT67}]
{\includegraphics[height=0.4\hsize,clip]{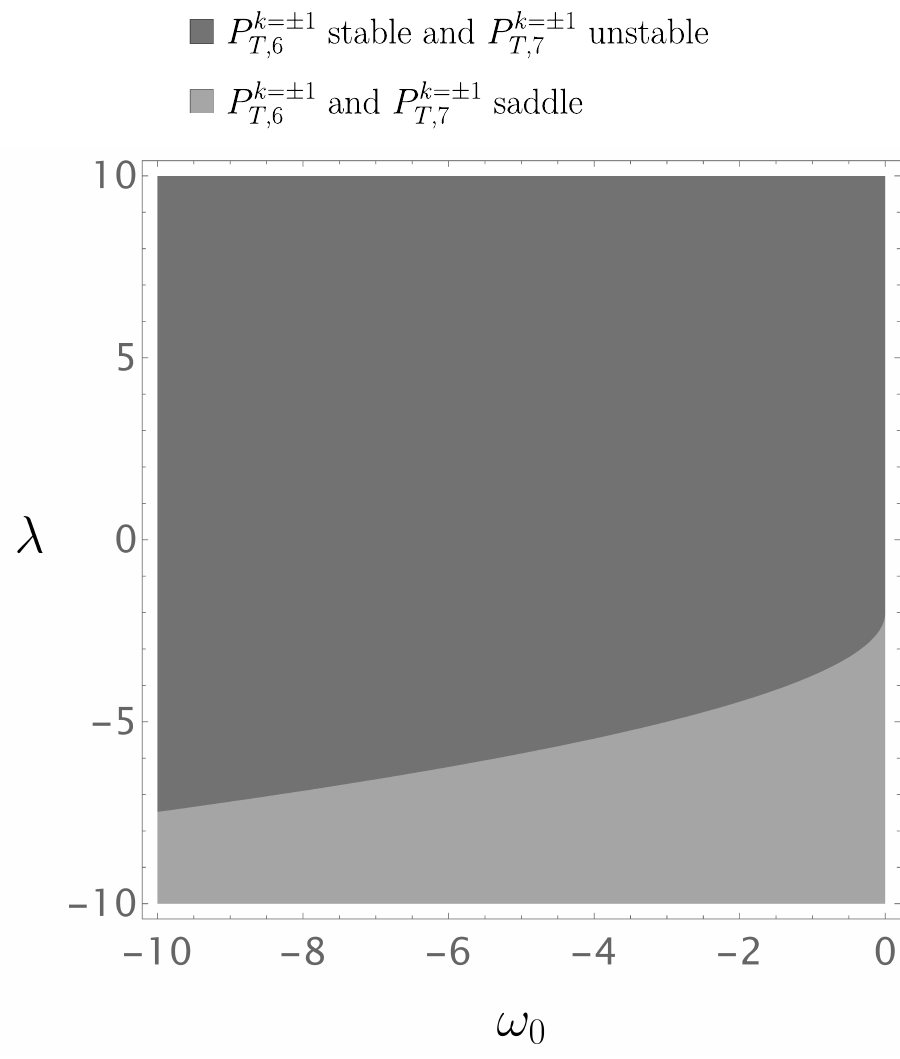}}
\subfigure[$\{\omega_0,\lambda\}$ region to identify the stability of $P_{T,8}^{k=1}$ and $P_{T,9}^{k=1}$.\label{PT89}]
{\includegraphics[height=0.4\hsize,clip]{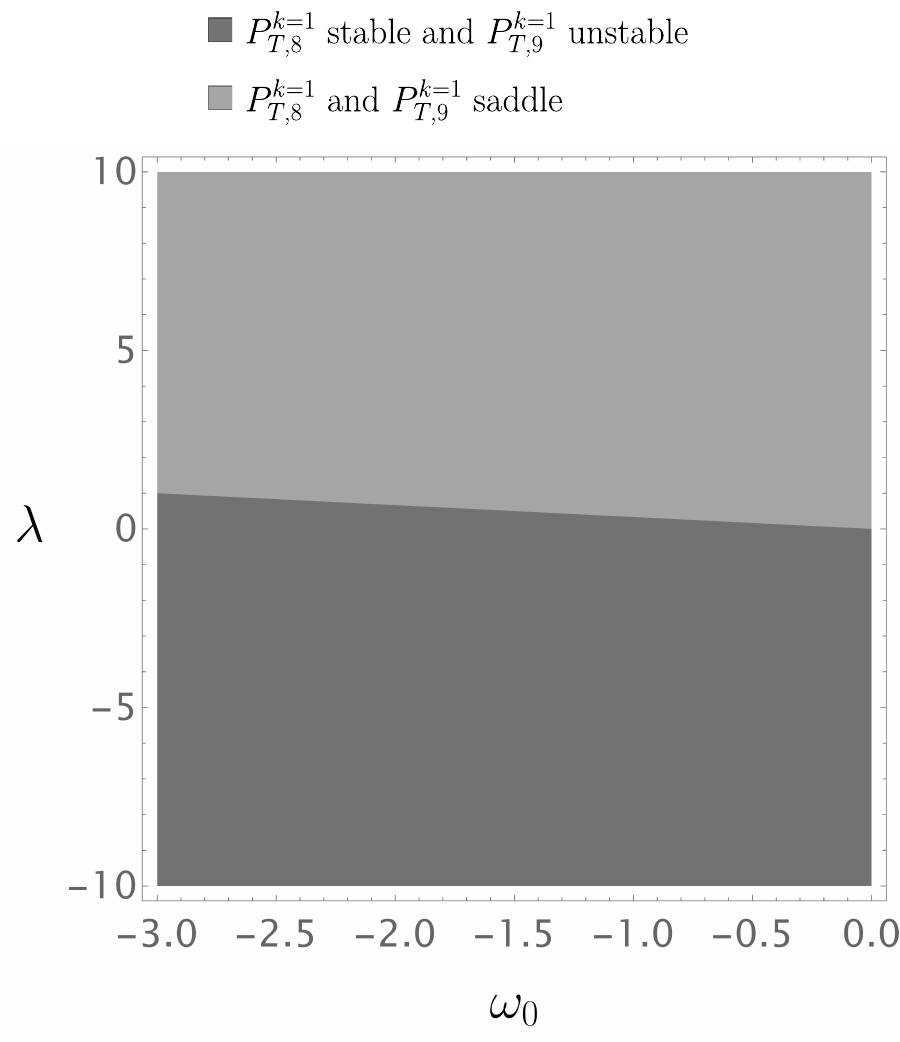}}
\subfigure[$\{\omega_0,\lambda\}$ region to identify the stability of $P_{T,8}^{k=-1}$ and $P_{T,9}^{k=-1}$.\label{PT89k-1}]
{\includegraphics[height=0.4\hsize,clip]{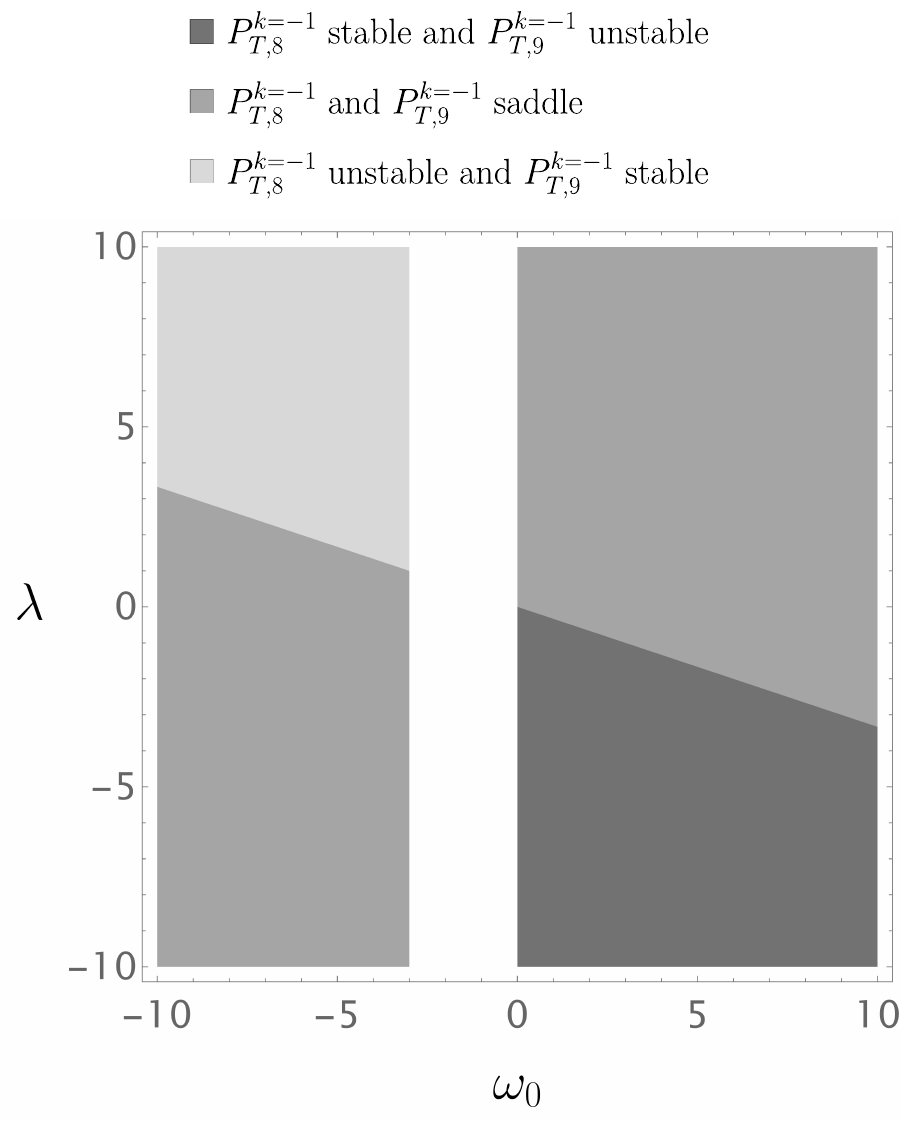}}
\caption{Free parameter regions to establish scalar field stability in teleparallel gravity for both closed and open universes.}\label{fig:stabcond2}
\end{figure*}

The differences between these two theoretical frameworks are manifest, as illustrated in Figs.~\ref{fig:stabcond1}–\ref{fig:stabcond2}. Nevertheless, a certain degree of degeneracy arises when the initial cosmological configuration is assumed to correspond to either a closed or an open universe. In particular, for the vast majority of critical points examined, the associated stability properties remain invariant with respect to the choice of spatial curvature.

For completeness, the stability regions corresponding to the two gravitational theories are reported below.

\end{document}